\documentclass[a4paper]{article}

\usepackage[utf8]{inputenc}
\usepackage[T1]{fontenc}
\usepackage[english]{babel}
\usepackage[protrusion=true,expansion=true]{microtype}
\usepackage{graphicx}
\usepackage{subcaption}
\usepackage{amsmath}
\usepackage{cite}
\usepackage{listings}
\usepackage{authblk}

\usepackage{textcomp, gensymb}
\usepackage{placeins}
\usepackage{hyperref}

\newcommand{\pt}{p_\mathrm{T}}
\newcommand{\kt}{k_\mathrm{T}}
\newcommand{\ktdurham}{\kt^\mathrm{Durham}}
\newcommand{\abseta}{|\eta|}
\newcommand{\PYTHIA}{\textsc{Pythia}}
\newcommand{\lambdaPV}{\lambda_{\mathrm{PV}}}
\newcommand*{\eV}[1][]{\mathrm{#1 e\kern-0.15ex V}}

\title{
Hunting for vampires and other unlikely forms of parity violation at the Large~Hadron~Collider
} 
\author[1]{Christopher~G.~Lester$^*$}
\author[2]{Radha~Mastandrea$^*$}
\author[1]{Daniel~Noel$^*$}
\author[1]{Rupert~Tombs%
\footnote{Authors are listed in alphabetical order.}}
\affil[1]{%
Cavendish~Laboratory,
University~of~Cambridge,
CB3~0HE,
United~Kingdom
}
\affil[2]{%
Department~of~Physics,
University~of~California,
Berkeley, 
CA~94720, 
USA
}
\date{13th July 2022}

\begin{document}

\maketitle

\section*{Abstract}

\noindent
Non-Standard-Model parity violation may be occurring in LHC collisions.
Any such violation would go unseen, however, as searches are for it are not currently performed. 
One barrier to searches for parity violation is the lack of model-independent methods sensitive to all of its forms.
We remove this barrier by demonstrating an effective and model-independent way to search for parity-violating physics at the LHC. 
The method is data-driven and makes no reference to any particular parity-violating model. 
Instead, it inspects data to construct sensitive parity-odd event variables (using machine learning tools), and uses these variables to test for parity asymmetry in independent data.
We demonstrate the efficacy of this method by testing it on data simulated from the Standard Model and from a non-standard parity-violating model.
This result enables the possibility of investigating a variety of previously unexplored forms of parity violation in particle physics.

Data and software are shared at \url{https://zenodo.org/record/6827724}.





\section{Introduction}

Sources of parity violation which are due to effects beyond the Standard Model have the potential to be observed in Large Hadron Collider data. 
Alas, there are many new and unexpected ways in which parity can be violated, and no continuous parity-odd event variable is sensitive to them all~\cite{lesterChiralMeasurements2021}. 
Furthermore, since there are not many commonly-studied%
\footnote{
See later comments that BSM parity-violation models, if they are to have observable signals in the context of this work, must abandon at least one of: (i) locality, (ii) Lorentz-invariance, or (iii) a basis in quantum field theory.  Vampires are not visible when mirrored~\cite{dracula:bram:stoker} and so presumably the laws of physics describing them derive their parity-violation from the loss of at least one of the above three properties!
}
theoretical models which predict forms of parity violation which could be visible at the LHC, there are few guides as to where and how to look for non-standard parity-violating signals. 
The absence of practical, general, \textit{and} scalable ways of performing such searches has been, until very recently, a significant barrier to performing them.\footnote{Only a single search,~\cite{Lester:2019bso}, has been performed on real LHC data so far! Furthermore,~\cite{Lester:2019bso} described many of its own choices as arbitrary and lacking in any strong theoretical motivation or generality. As a result, its null result is of limited value, even if there is value in the existence of the work as a proof-of-principle that data-data searches for non-standard sources of parity-violation can be performed.}
 
The primary purpose of this paper is to demonstrate that the data-driven parity-violation search strategies proposed by~\cite{lester2021stressed} and~\cite{tombs2021which} are, in fact, capable of removing the above barrier to realistic searches for parity violation in particle physics.
Our work is necessary because the methods proposed in 
\cite{lester2021stressed} are demonstrated only with toy examples from outside of particle physics, while those in~\cite{tombs2021which} use neither real nor simulated particle physics datasets holding signatures of parity-violation. 
The present work will demonstrate the viability of the new approaches by showing sensitivity to a range of parameters within an example parity-violating model, which is described within the Lorentz-invariance violating framework of the minimal Standard Model Extension~\cite{LIV:Main:Colladay:1998fq}. For this model, we generate Monte Carlo simulated events assuming a barrel-shaped LHC detector such as ATLAS, CMS, or ALICE.

Specifically,~\cite{lester2021stressed} points out that a clear signal for parity violation can be obtained from a parity-odd event variable if it is seen to have an asymmetric distribution.
An optimization using one part of a dataset can be used to create such a variable, while another part of the dataset can be used to test for the suspected asymmetry in the distribution of that variable.  
For the reasons given in~\cite{lester2021stressed}, this method presents a very general, flexible, and scalable way of maintaining sensitivity to all potential source of new physics which are able to generate events which are distinguishable (in distribution) from their parity-flipped partners --- that is to say any models with `manifestly’ parity violating signatures. 
Whether any given source is observable would, of course, always depend on the sizes of systematic uncertainties and the amount of data taken.
It would also depend on whether the particular choice of parity-odd function(s) that someone implementing~\cite{lester2021stressed} had decided to use were sensitive to the form of parity violation which the given source induces in its events. 
However, those are questions of implementation rather than intrinsic limitations.
The generality of the method itself is an unavoidable consequence of the fact that it is, in effect, simply a codification of what it means for a model to have `manifestly’ parity violating consequences. 
In this sense, electing to use~\cite{lester2021stressed} does not place limits on what can be seen.

Known inefficiencies and acceptance effects, which act to filter data from observation in the detector, can also be easily corrected for in the manner described in~\cite{tombs2021which} if they are well understood. 
Such filtering effects are not, however, included in any detector models in this work.

Our approach can be viewed as an adoption of standard practice from the field of machine learning: 
in training, a general-purpose `machine' generates hypotheses about the structure of the data distribution.
Then in testing, those hypotheses are tested on independent data.
By doing this with parity-odd algorithms, 
success in the testing phase becomes evidence for parity violation.   

Various end-stage techniques could be used to assess the strength of that evidence.
To follow existing practice for LHC searches, the parity-odd variables can be histogrammed and interpreted with conventional statistical methods. 
This approach, which we illustrate in Appendix~\ref{app:interpretations}, allows one to implement systematic uncertainties at testing time as histogram variations that allow weakly broken parity symmetry

The nature of parity violation and its possible visibility at the LHC is discussed in Section~\ref{sec:pv_at_the_lhc}.
Parity-violating physics in an example quantum field theory, and the simulation of its effects, are described in Section~\ref{sec:pv_physics}.
Machine learning implementations of parity-odd, symmetry-invariant event variables are presented in
Section~\ref{sec:methods}.
The varied successes of these variables in detecting parity violation in simulated data are displayed in Section~\ref{sec:results}.

\section{Parity violation at the LHC}
\label{sec:pv_at_the_lhc}

It is experimentally possible to see parity violation in particle physics data.
Whether it \emph{will} be seen depends both on whether nature produces parity violating effects
and whether the appropriate data analysis is performed.

Particles and particle physics data are typically described as being sampled from a differential cross-section $\sigma(x)$,
where $x$ is some representation of those data.
We would like to test whether $\sigma(x)$ violates parity symmetry;
that is, we want to answer the question ``is $\sigma(x)$ different from $\sigma(\mathrm{P}x)$?'' for the parity operator $\mathrm{P}$.
However, we have finite data. 
With finite data, asymmetries may be too subtle to notice,
and with continuous distributions asymmetries may hide in intricate patterns which we as analysts would never think to check.
We can, however, take opportunities to disprove parity symmetry if nature happens to give clear signals.

We can maintain sensitivity to all forms of parity violation by considering only parity-odd event variables~\cite{lesterChiralMeasurements2021}.%
\footnote{%
This statement is precisely formulated in Theorem~2.26 and summarized in Section~2.9 of~\cite{lesterChiralMeasurements2021}.
Essentially, the absolute value of a parity-odd variable is parity-even, and all parity-asymmetric variables can be 
constructed as functions of even and odd parts.
}
To test parity symmetry, we can therefore take any parity-odd event variable $f(x)$, 
for which $f(x) = -f(\mathrm{P}x)$, and compare the positive and negative halves of its distribution under $\sigma(x)$.
If parity symmetry is respected by $\sigma(x)$, then events and their parity-flipped versions must all be produced at equal rates.
Equivalently, there must be equal event rates at all $\pm f$ points.
However, if an experiment sees an asymmetric distribution in $f(x)$, for example by histogramming many data, then parity symmetry must be violated in $\sigma(x)$. 

Hadron-hadron collisions, such as those currently produced at the LHC, have many symmetries.
Two identical, unpolarized beams are collided head-on (to a good approximation),
and, since physical space is known to be isotropic (to a very good approximation),
the scattering processes must be invariant to all rotations which do not change the beams.
These rotations include both those about the beam axis (by any angle), 
and those which swap the two beams by rotating $180^\circ$ about a perpendicular axis.
Furthermore, data often comprise unordered sets of four-momenta for otherwise-identical particles;
although these always take some conventional order in practice,
nature does not care about any ordering we assign.

When analysing such symmetric data, it is helpful to be blind to transformations under their symmetries;
otherwise, the handling of symmetrically equivalent, yet differently represented data
introduces unnecessary complexity.
This is one benefit of rotation-invariant event variables at the LHC,
such as those based on masses or transverse momenta $\pt$, 
and to collapsing the permutation symmetry of particles by $\pt$-ordering their representations.

Furthermore, our aim is only to test for violation of parity symmetry, 
and not for violations of other symmetries such as spacetime isotropy.
Blinding our analysis to violations of other symmetries, therefore, 
helps the method to focus more precisely on the task of testing parity violation itself.
For these reasons, when constructing parity-odd event variables in this paper,
we choose to make them also rotation-, beam-swap- and permutation-invariant.
This invariance is achieved by design in both data representations and function architectures.

\begin{figure}[t]
\centering
\includegraphics[width=0.999\textwidth]{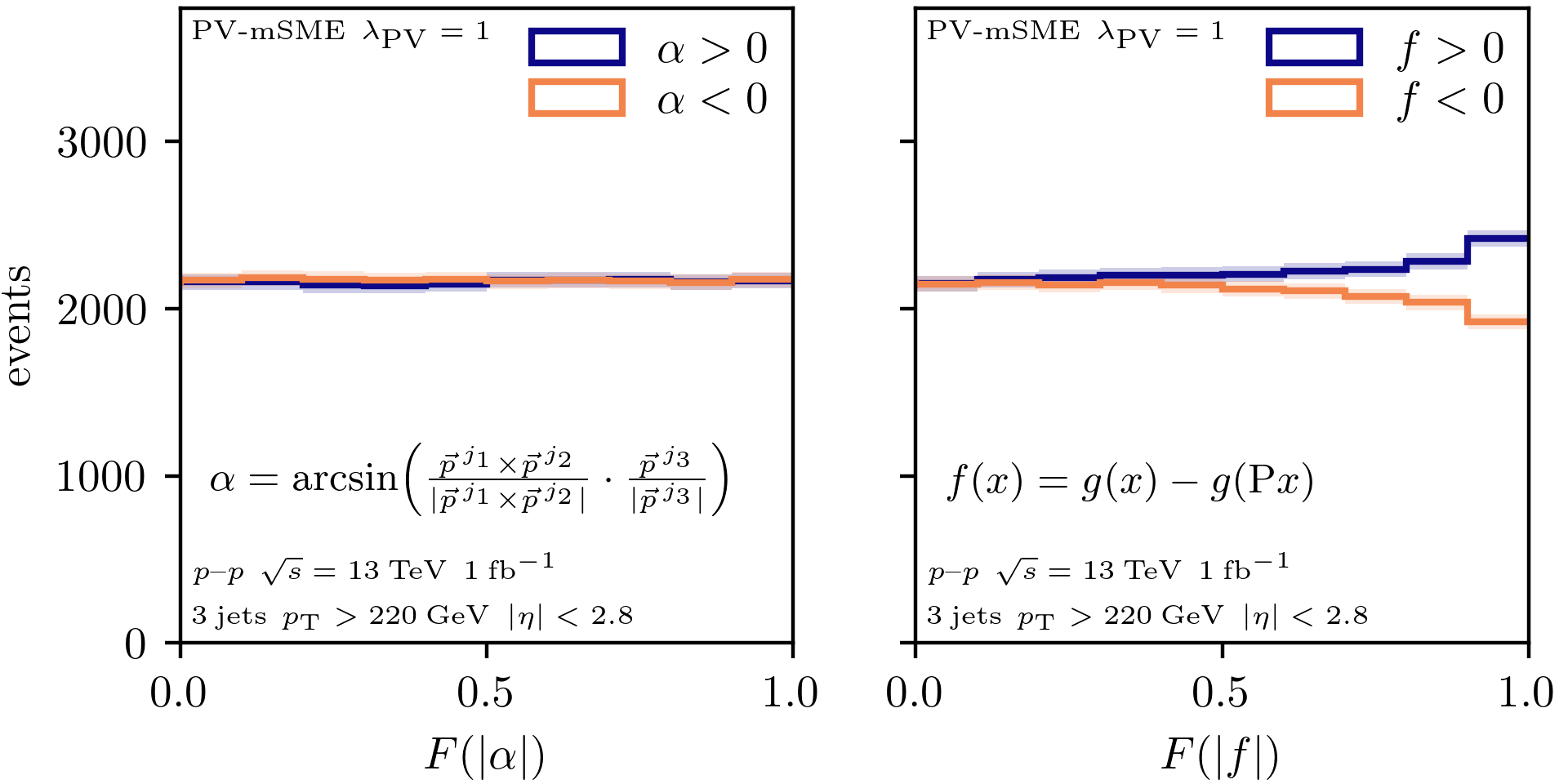}
\caption{
Distributions of two different parity-odd variables for the same parity-violating model of physics:
(left) $\alpha$ as defined in Equation~\ref{eq:cms_alpha} and (right) a parity-odd neural network looking at reconstructed jet momenta. 
Samples are normalized to the Standard Model cross-section at $1~\mathrm{fb}^{-1}$ of LHC 
$\sqrt{s}=13~\eV[T]$ proton-proton collision data.
The histograms contain only data from an independent testing set containing $20\%$ of the total yield.
Each function $F(\cdot)$ is a transformation fitted to training data to make each distribution close to uniform.
Error bars are of size $\pm\sqrt{n}$ to show the size of statistical uncertainty at the chosen luminosity.
}
\label{fig:plots:hist_both_alpha_net_transformed_linear}
\end{figure}

One example of a parity-odd, rotation-, beam-swap- and permutation-invariant event variable is
\begin{equation}
\label{eq:cms_alpha}
\alpha
= \arcsin\!
\left(
    \frac{
        \vec p^{\,j_1} \times \vec p^{\,j_2}
    }{
        |\vec p^{\,j_1} \times \vec p^{\,j_2}|
    }
    \cdot
    \frac{
        \vec p^{\,j_3}
    }{
        |\vec p^{\,j_3}|
    }
\right)
\end{equation}
in which $\vec p^{\,j_i}$ is the momentum of the $i\textrm{th}$ hardest jet. 
This $\alpha$ variable was proposed and used in~\cite{Lester:2019bso} to demonstrate that parity violation can be searched for in events with three or more jets.
It has rotation invariance from the basis-independence of dot and cross products, permutation invariance from the $\pt$-ordering of jets, and parity-oddness since the three momentum vectors each pick up a factor of $-1$ under the parity flip.
It also has, however, ``no claims of generality or optimality''~\cite{Lester:2019bso}.
Indeed, Figure~\ref{fig:plots:hist_both_alpha_net_transformed_linear} illustrates that it
does not have good sensitivity to parity violation in a sample of simulated data, described below in Section~\ref{sec:pv_physics},
while an observable $f$, created using the methods of \emph{this} paper, does.

Unfortunately, parity violation in \emph{data} does not necessarily imply parity violation in \emph{nature}.
Data result from compositions of natural physical processes with the actions of experimental hardware and software, 
and any of these parts could introduce asymmetries.
Fortunately, any detector effect that can be anticipated and that acts as a filter (reducing data rates in certain selections),
can be removed (i.e.,~the possibility of false positive excluded) using the method of~\cite{tombs2021which}.

Nonetheless, unknown unknowns could still cause problems.
A positive signal for parity violation therefore provides either: (a) evidence for parity violation in \textit{nature}, or (b) evidence for detector effects and/or calibrations that need to be improved. 
Both possibilities are interesting and valuable, since one tells us something new about nature, while the other tells us how to improve our detector calibrations which will benefit other analyses (even those not interested in symmetry violations) being performed by the same detector.

\section{Parity violating physics}
\label{sec:pv_physics}

\subsection{PV-mSME (Parity Violating minimal Standard Model Extension)}
Nature may, one day, not be described by a quantum field theory (QFT). At present, however, the majority of existing theories are QFTs, and so we choose to illustrate our method with one.  
Locally Lorentz invariant QFTs which violate parity\footnote{To be precise, one should say: `Locally Lorentz invariant QFTs for which parity violation is theoretically observable in data derived from collisions of unpolarized particles in the case that the final-state event data recorded contain only four-momenta which are not decorated with additional labels (such as helicities or flavours).}  must necessarily violate charge-parity (CP) symmetry for the reason explained in Footnote~1 of~\cite{Lester:2019bso}. 
Although one could therefore choose to illustrate our method by searching for parity violation in a CP-violating model, doing so is unlikely to be competitive with pre-existing methods as tests of CP-violation are highly developed with a mature field; in any case, the amount of CP-violation in the Standard Model is known to be small. 

We therefore choose to use a Lorentz-violating extension of the Standard Model, namely the `minimal Standard Model Extension' (mSME) of~\cite{LIV:Main:Colladay:1998fq}. 
We do not work with the full generality of the mSME, but instead use only a small subset of its parity-violating sector, which we name the `PV-mSME'. 
Within the PV-mSME, we control the strength of parity violation with a single real parameter $\lambda_{\mathrm{PV}}$, which is described below.

The mSME makes many extensions to the Standard Model, among which is a term 
$\mathcal{L}^\textrm{CPT-even}_\mathrm{quark}$ (defined in Equation~11 of~\cite{LIV:Main:Colladay:1998fq})
which modifies the quark-quark-gluon Feynman Rule.
We choose to use only this extension from the mSME, in order to generate parity violation in strongly produced multi-jet events, which are attractive to study since they have large cross-sections at the LHC and are not expected to show parity violation in the Standard Model.
Multi-jet events were also previously studied parity violation in~\cite{Lester:2019bso}.
Furthermore, we choose couplings to switch on only its axial vector part with the intention of violating parity in interference with the Standard Model vector interactions.

The Standard Model quark-quark-gluon vertex from takes the form
\begin{align}
f_\nu^\textrm{SM}
&= -i g_s \frac{\lambda_{ij}^a}{2}
\gamma_\nu,
\intertext{
to which the PV-mSME adds an axial vector part shaped by a coupling matrix $(c_{A})_{\mu\nu}$,
}
f_\nu^\textrm{PV-mSME}
&= -i g_s \frac{\lambda_{ij}^a}{2}
\label{eq:pv-sme-qqg-vertex}
\left(
\gamma_\nu
+
\gamma^\mu
(c_{A})_{\mu\nu}\gamma^5
\lambda_\textrm{PV}
\right)
,
\end{align}
in which the strength parameter $\lambdaPV$ is varied to control the magnitude of parity violation in the model.
Motivated by reasons given in Appendix~\ref{app:pvsme_details}, we define $(c_{A})_{\mu\nu}$ to be
\begin{equation}
\label{eq:camunu_pv_mSME}
(c_{A})_{\mu\nu}
= 
\begin{pmatrix}
    0 & 0 & 0 & -1 \\
    0 & 0 & -1 & 0 \\
    0 & 1 & 0 & 0 \\
    1 & 0 & 0 & 0 \\
\end{pmatrix}
\end{equation}
such that the value $\lambdaPV = 1$ results in
PV-mSME terms having similar magnitudes to Standard Model terms in matrix elements.
Clearly, the Standard Model is recovered for $\lambdaPV = 0$. 

Although we are not aware of direct constraints on this exact usage of the mSME, 
data from astrophysics and deep inelastic scattering have been used to limit the
couplings in this mSME quark sector to very small values of at most $10^{-4}$ in 
magnitude~\cite{LIV_datatables,KOSTELECKY2017272}.
Large $\lambdaPV$ values would therefore be excluded by current data if tested.

Unlike the common background-plus-signal scenario,
PV-mSME cross-sections do not add linearly since they interfere with the Standard Model,
nor do they scale linearly with the coefficient $\lambdaPV$.
Indeed, squared matrix elements computed from a diagram with $n$ quark-quark-gluon vertices will, in general, contain terms with all powers from $(\lambdaPV)^0$ up to and including $(\lambdaPV)^{2n}$.

Couplings in the mSME break spatial isotropy; they induce special spatial directions in which particles have differing behaviours.
Having constant couplings, as we do from Equation~\ref{eq:camunu_pv_mSME}, 
therefore ignores rotations of the detector through space.
Detectors at the LHC do, in fact, rotate --- with Earth through the day, and about the sun throughout the year, and so on.
Rotations could be controlled for, for example, by selecting data from only times when the detector is aligned within a small solid angle of a special direction, or by relocating the LHC to a non-rotating spaceship. 
Neither solution is practical, since the first slashes the event rate, and the second is beyond CERN's budget.
Alternatively, one could employ a different, fully rotation-aware method of data generation and analysis, which we discuss
further in Appendix~\ref{app:rotation}.

Faced with the decision of whether to account for the Earth's rotation, we elect to make no correction and choose to leave our PV-mSME coupling matrixes constant in the laboratory frame.  
In effect, this means we are working with a model in which Lorentz violating couplings are `dragged around' by Earth, or where Earth does not spin.
We make this choice, despite it being unphysical, because making those corrections would complicate the analysis without adding anything useful. 
Moreover, if we were to correct for Earth's rotations we would be building assumptions into our analysis that are not explicitly related to parity, and we do not wish to do so.

Alternatively, our choice of constant couplings could be implemented by selecting data collected at dates 
and times when the detector aligned with a special direction.
In reality, direction dependence might be handled better by encoding the detector's alignment in additional event variables, so that algorithms can learn alignment-dependent parity violation.
We demonstrate a simple form of this rotation-encoding in Appendix~\ref{app:rotation}, and find that weaker signals of parity violation can still be observed in the PV-mSME on a rotating planet.

Despite this unphysical use of the mSME, our PV-mSME remains a parity violating quantum field theory,
and provides a parity violating model with which we can demonstrate our methods

\subsection{Simulation}
A complete simulation pipeline is performed to produce realistic simulations of proton-proton collisions scattering to jets in the PV-mSME with various settings of $\lambdaPV$ between $0$ and $1$.
This pipeline comprises several standard steps:
partons are first simulated in a hard scatter of protons to three or four gluons or light quarks,\footnote{Note that a $2\rightarrow2$ scattering process could never be parity violating, as all participating 4-momenta lie in a plane.}
then dressed with an underlying event, parton shower and pileup overlay,
and hadronized into observable particles.
A detector response to these energetic particles is simulated and processed into event variables.

We introduce PV-mSME effects through matrix elements in the hard scatter only;
any changes that it should make to parton distributions, showering, hadronization, and detector physics are ignored in this work.

Triggering is modelled by selecting only events with at least three central jets with $\pt > 220~\eV[G]$;
this selection is designed to approximate the efficiency plateau of the three-jet trigger in a general-purpose LHC detector.

Details of the simulation software and settings are given in Appendix~\ref{app:simulation_details}.

\begin{figure}[t]
\centering
\includegraphics[width=0.49\textwidth]{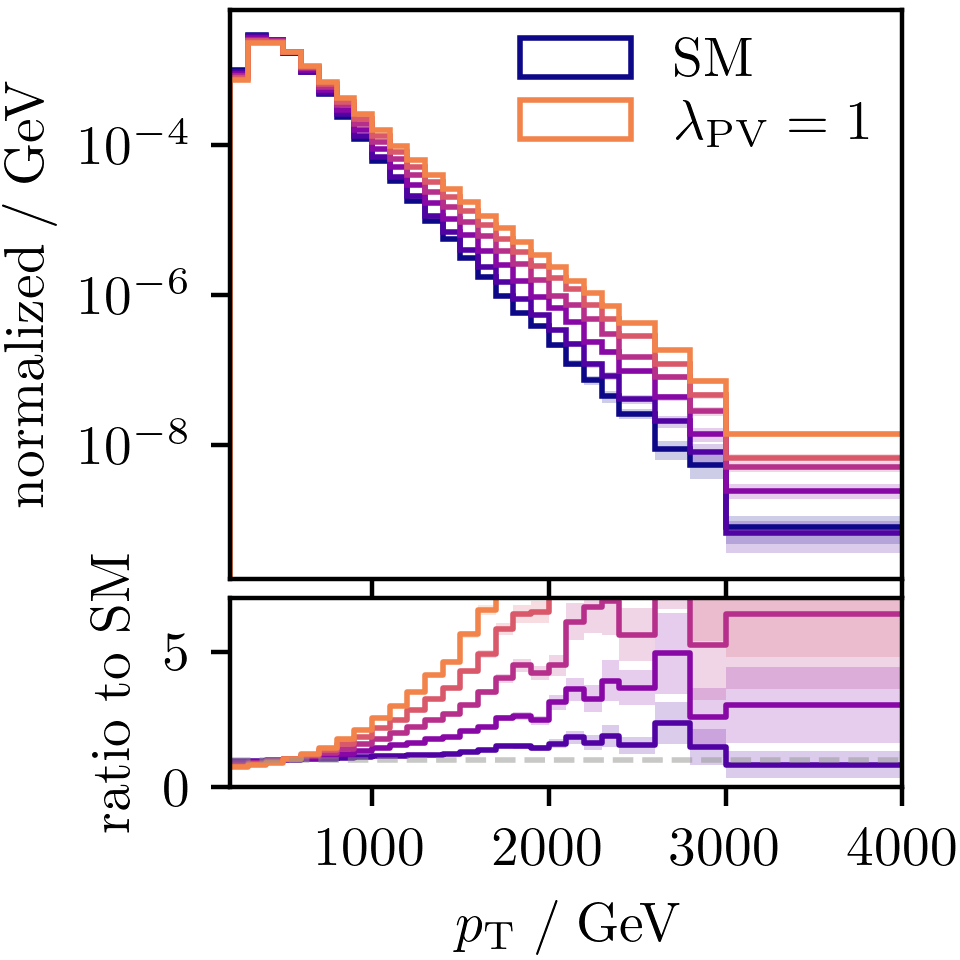}
\includegraphics[width=0.49\textwidth]{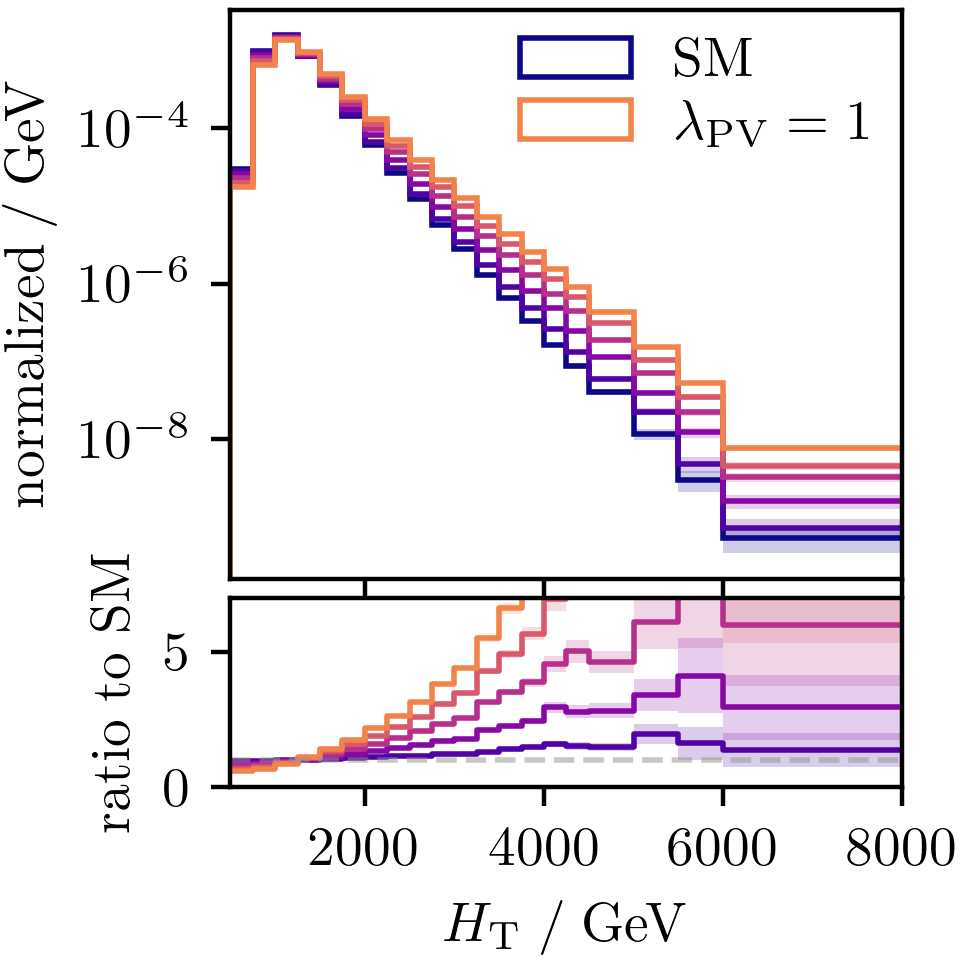}
\caption{
Distributions of kinematic variables of reconstructed jets with varied $\lambdaPV$ in the PV-mSME.
(left) Transverse momentum of the hardest jet and (right) the scalar sum of transverse momenta of the hardest four jets in the event.
The five coloured lines linearly interpolate from the Standard Model ($\lambdaPV = 0$) to $\lambdaPV$ = 1.
}
\label{fig:plots:ht_pt_reco}
\end{figure}

Although we primarily target events with three jets, 
we include four-parton processes at truth level as a sub-leading effect
to increase physical accuracy.
At reconstruction level after kinematic selections, the Standard Model cross-section 
of this multi-jet process is
$0.22~\mathrm{nb}$
of which $0.7\%$
is attributed to four-parton processes at truth level.
This increases with $\lambdaPV$ to 
$0.31~\mathrm{nb}~(9\%)$ for $\lambdaPV = 0.3$
and
$2.6~\mathrm{nb}~(27\%)$ for $\lambdaPV = 1$.
Evidently, large $\lambdaPV$ drastically increases production rates and jet multiplicities.
Kinematic shapes also change significantly, as demonstrated in Figure~\ref{fig:plots:ht_pt_reco}.
Our interest, however, is in the parity violation of these kinematic distributions, independent of other effects.
We therefore normalize the PV-mSME distributions to the Standard Model cross-section wherever relevant.

\subsubsection{Data formats}
Simulation results are processed into two kinds of event variables for analysis;
the first is standard, using estimated four-momenta of jets,
while the second bypasses jet reconstruction to produce low-resolution images of energy deposits in the (cylindrical) calorimeters, projected into the $\eta\textrm{--}\phi$ plane, where $\eta$ is the pseudorapidity and $\phi$ is an angle about the beam pipe.

\begin{figure}[tp]
\centering
\includegraphics[width=0.999\textwidth]{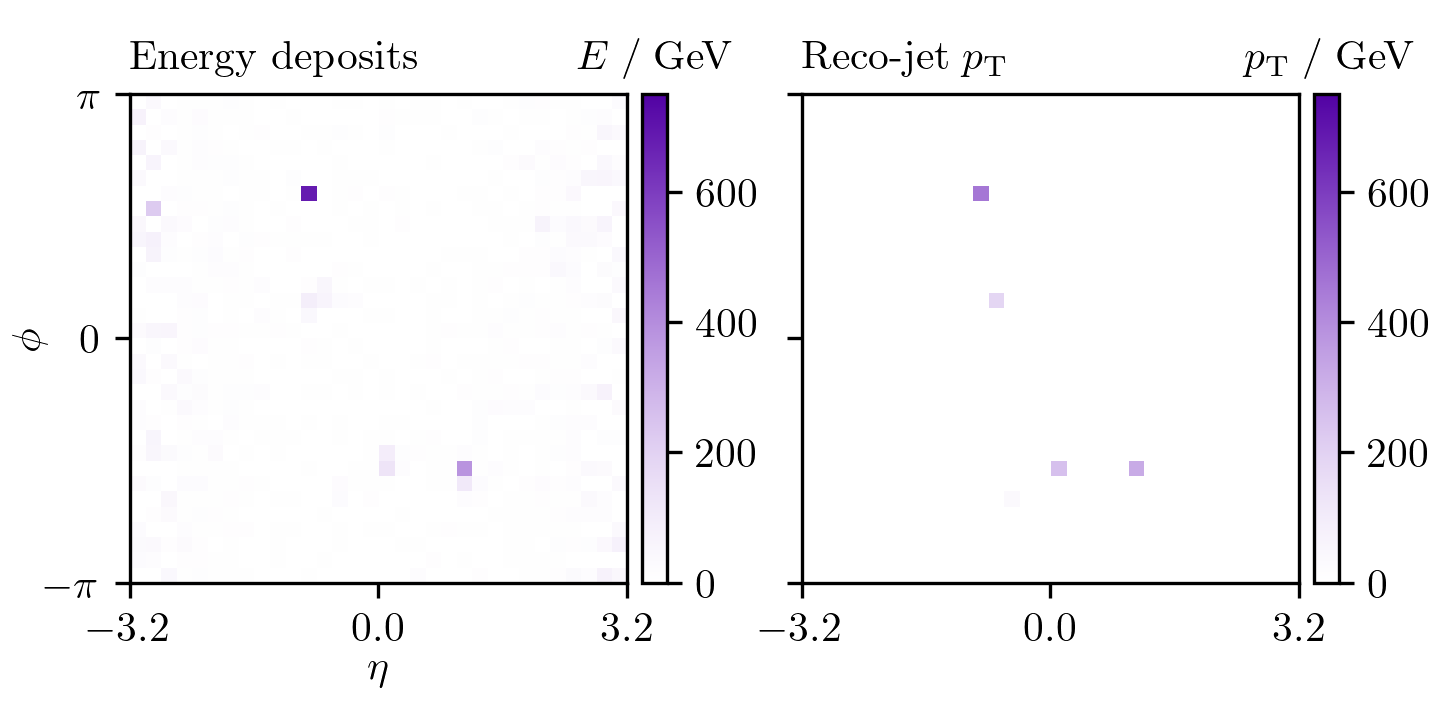}
\caption{Energy deposits in the calorimeter (left) and the $p_T$ of reconstructed jets (right) for an example Standard Model event.}
\label{fig:energydeposit_jet_image}
\end{figure}

These images consist of $32\times 32$ pixels that cover the entire $2\pi$ range in $\phi$ and $\abseta < 3.2$, which corresponds to the extent of the ATLAS calorimeters, excluding the forward calorimeter~\cite{atlasdetector}.
Images are constructed either directly from calorimeter energy deposits or indirectly from jets by histogramming their transverse momenta in two-dimensional $\eta\textrm{--}\phi$ histograms, where every jet with $\pt > 30~\eV[G]$ and $\abseta < 2.8$ is included.
Images of an example event are shown in Figure~\ref{fig:energydeposit_jet_image}.


To assess both how sensitivity degrades with noise and how well machine learning tools perform with different data formats, we use three representations of the simulated data.
These are:
\begin{itemize}
    \item `truth-jet': the true momenta of partons from the hard scatter,
    \item `reco-jet': reconstructed momenta of the four hardest jets, and
    \item `calo-image': images of calorimeter energy deposits.
\end{itemize}
The next section describes how of these data representations is used with each machine learning model.

\section{Methods}
\label{sec:methods}
As stated in Section~\ref{sec:pv_at_the_lhc}, we can search for parity-violating physics using a parity-odd event variable $f(x)$. 
Such a variable can easily be constructed from an arbitrary function $g(x)$ since
\begin{equation}
\label{eq:parity_odd_fun}
f(x) = g(x) - g(\mathrm{P} x)
\end{equation}
is parity odd.\footnote{%
Numerically, this works for all numbers except signed zeros, 
since if $g(x) - g(\mathrm{P} x) = +0.0$, 
then $g(\mathrm{P} x) - g(x) = +0.0$, not $-0.0$, in standard floating-point arithmetic.
Zeros of $f(x)$ must therefore be excluded from parity-odd interpretations.}
All of our models use this construction.
Machine learning algorithms present practical, general, and scalable ways to assign functional forms to $g(x)$,
and our method does not depend on the exact algorithm chosen.
To demonstrate that, we use boosted decision trees (BDTs), neural networks (NNs), and convolutional neural networks (CNNs) to build the functions $g(x)$.

The BDT and NN approaches use vectors of jet momentum features derived from truth-jet and reco-jet data, whereas the CNN approach uses either calo-image data or jet features transformed into images as described in Section~\ref{sec:invariant_images}.

\FloatBarrier
\subsection{Symmetries}

We ensure that rotational and permutation symmetries are preserved in our parity-odd event variables,
for reasons discussed in Section~\ref{sec:pv_at_the_lhc}.
This section describes how those symmetries are enforced for the various models and data formats.

\subsubsection{Invariant jets}
\label{sec:invariant_jets}
Jet permutation symmetry is ensured by sorting objects from hardest to softest in $\pt$.
Invariance to rotations (along and about the beam axis) is ensured 
by rotating all momenta to an orthonormal basis 
$\{\hat x, \hat y, \hat z\}$ based on the hardest jet momentum $\vec p^{\,j_1}$;
$\hat x$ is taken in the transverse direction of $\vec p^{\,j_1}$,
$\hat z$ is taken along the beam axis with the sign of its longitudinal momentum $\vec p^{\,j_1}_z$,
and the third basis vector is taken as their cross product; $\hat y = \hat z \times \hat x$. 

Mutual transformation of the hardest jet with all other momenta under rotations
means that the projection of momenta onto this basis is unchanged under rotations
(up to numerical precision).
Defining each element of momentum in this new rotation-invariant basis as 
$\vec q^{\,j_i}_a = \vec p^{\,j_i} \cdot \hat a$ for $a \in \{x, y, z\}$,
it is clear that 
$\vec q^{\,j_1}_y = \vec p^{\,j_1} \cdot \hat y = 0$.
That is, the $y$ component of the hardest jet is zero by definition and not useful, so we discard it.

Parity flipping momenta in laboratory coordinates leads only to changing the signs of $\vec q^{\,j_i}_y$ components, since under parity
\begin{align}
\hat x &\rightarrow -\hat x~,\\
\hat z &\rightarrow -\hat z~,\text{ but}\\
\hat y &\rightarrow -\hat z \times -\hat x = +\hat y~;
\end{align}
the $\vec q^{\,j_i}_x$ and $\vec q^{\,j_i}_z$ components are parity-even, and $\vec q^{\,j_i}_y$ components are parity-odd.
Both parity-odd and parity-even event variables can be useful in the search for parity violation;
although parity-flipped events differ only in their parity-odd parts, the even parts
are context which can be required to separate regions in which the parity flipped versions
occur at different rates~\cite{lester2021lorentz}.

\subsubsection{Invariant images}
\label{sec:invariant_images}
Invariance to rotations (about and along the beam axis) is built into the CNN structure.
For invariance to $\phi$-rotations, note that a rotational symmetry in $\phi$ corresponds to a wrap-around translational symmetry in $\phi$ when considering the $\eta\textrm{--}\phi$ plane.

Each image is padded cyclically around the $\phi$ axis before applying each convolutional layer; this maintains equivariance.
Invariance to $\phi$ rotations is achieved by performing a max-pool operation over the entire $\phi$-axis after the convolutions.
This pooling selects the largest pixel value for each slice in $\phi$, which is clearly invariant to  $\phi$ rotations of those pixels.
This design gives perfect invariance to discrete rotations of $2\pi/32$, but the CNN is not perfectly blind to other rotations prior to pixelization; this could incur some costs for the reasons discussed in Section~\ref{sec:pv_at_the_lhc} which encourage invariance.

For invariance to discrete beam-flip rotations, $R_{\pi}$ (rotation by $180^\circ$ about an axis perpendicular to the beams),
we construct $R_{\pi}$-invariant $g(x)$ functions (of Equation~\ref{eq:parity_odd_fun}) as $g(x) = h(x) + h(R_{\pi} x)$, similarly to how parity-oddness is achieved. 
Then, 
\begin{equation}
\label{eq:parity_odd_cnn_output_two}
f(x) = [h(x) + h(R_{\pi}x)] - [h(\mathrm{P}x) + h(R_{\pi}\mathrm{P}x)]
,
\end{equation}
and we learn a form for $h(x)$ with a CNN.
This maintains the parity-odd property $f(x) = -f(\mathrm{P}x)$ and introduces the beam-flip symmetry $f(x) = f(R_{\pi}x)$, since $R_{\pi}R_{\pi}x = x$.

The parity flip operator transforms $\eta \rightarrow -\eta$ and $\phi \rightarrow \phi + \pi \mod  2\pi$. With the rotationally invariant CNN, $\phi$ changes are inconsequential, so the parity flip operator is just a negation of $\eta$, or equivalently a mirror image through the line $\eta = 0$.

\begin{figure}[tp]
\centering
\begin{subfigure}[b]{0.49\textwidth}
    \centering
    \includegraphics[width=\textwidth]{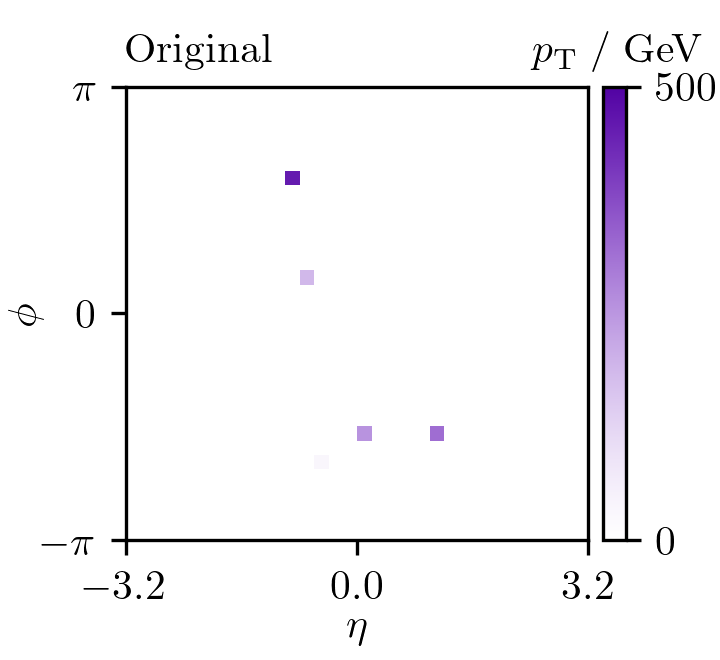}
    \caption{}
\end{subfigure}
\begin{subfigure}[b]{0.49\textwidth}
    \centering
    \includegraphics[width=\textwidth]{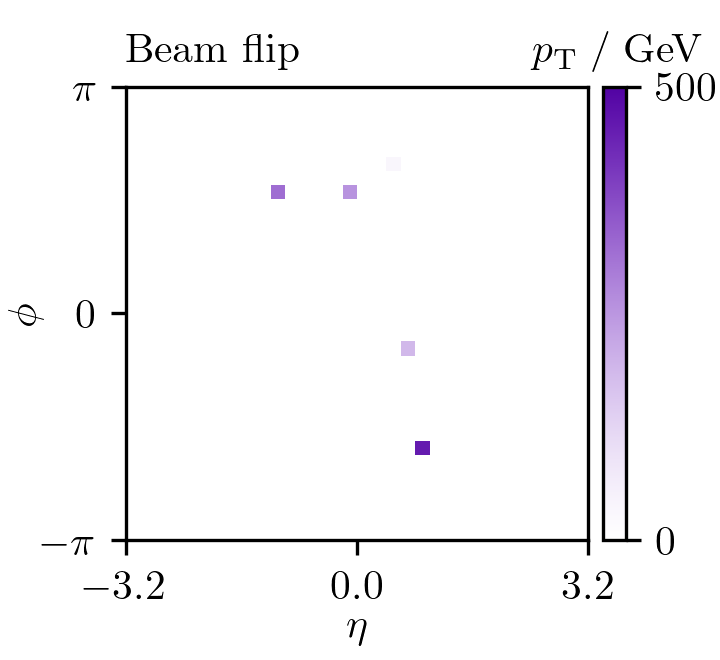}
    \caption{}
\end{subfigure}
\begin{subfigure}[b]{0.49\textwidth}
    \centering
    \includegraphics[width=\textwidth]{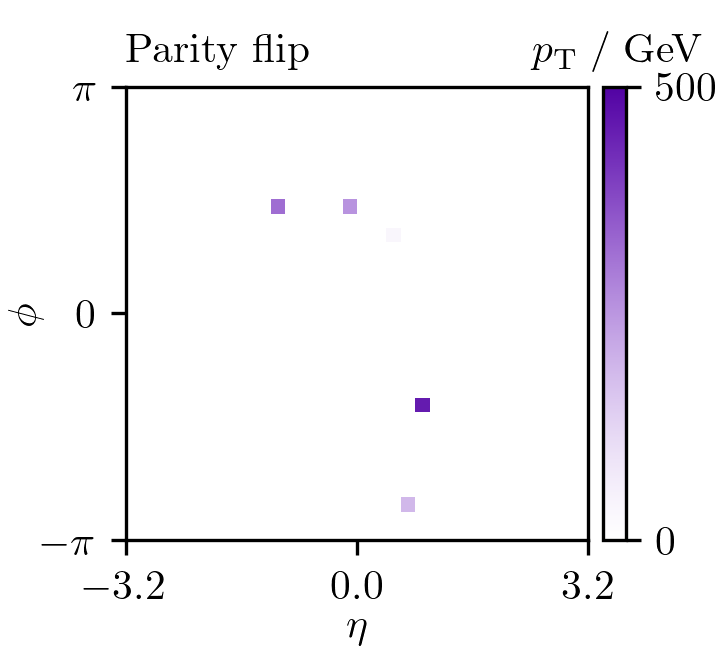}
    \caption{}
\end{subfigure}
\begin{subfigure}[b]{0.49\textwidth}
    \centering
    \includegraphics[width=\textwidth]{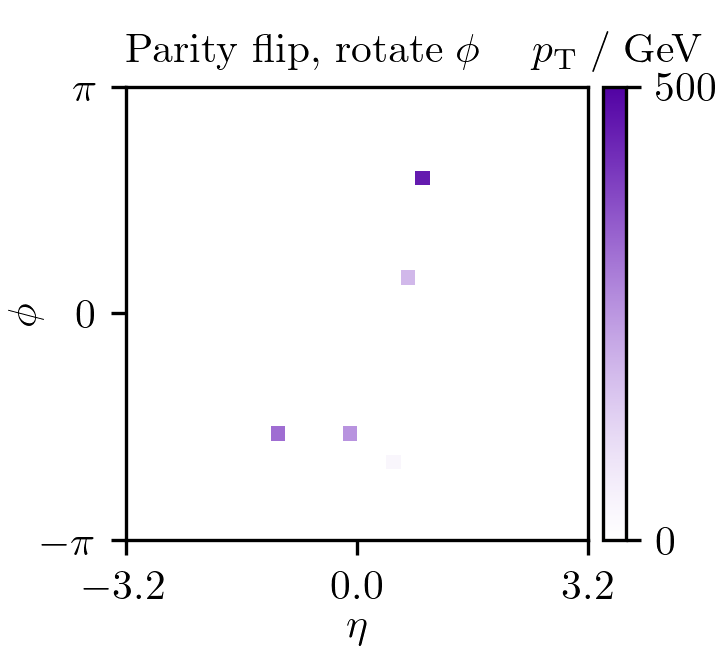}
    \caption{}
\end{subfigure}
\caption{Histogrammed $p_T$ of reconstructed jets for a Standard Model event projected in the $\eta\textrm{--}\phi$ plane. (a) the original event image (b) rotated $180\degree$ about the $x$-axis (c) parity flipped (d) parity flipped and rotated by $\pi/2$ in $\phi$. The net output $f(x)$ is $-0.132$ for (a) and (b), and $0.132$ for (c) and (d); the symmetries we build into the network can be identified here.
}
\label{fig:cnn_symmetries_built_in}
\end{figure}

An example event image is shown in Figure~\ref{fig:cnn_symmetries_built_in}, along with its transformed copies. 
From the network outputs reported in the caption, we see that the network is indeed parity-odd whilst being symmetric to beam-flips. 
In addition, by comparing the outputs with a translation in $\phi$, we can see that the network is invariant to translations in $\phi$. 
This illustrates how a parity flip on an image in the $\eta\textrm{--}\phi$ plane is equivalent to a flip in $\eta$.

\subsection{Training}
Models are trained from gradients of the `which is real?' objective function introduced in~\cite{tombs2021which},
which is a special case of the standard `cross-entropy' (negative binomial log-likelihood) loss function~\cite{MurphyKevinP.2012Mlap} for binary classification.
Its class labels are the `real--fake' or `fake--real' orderings of an $\{x, \mathrm{P}x\}$ or $\{\mathrm{P}x, x\}$ pair,
where each pair contains both a real observed event $x$ and its parity-flipped `fake' representation $\mathrm{P}x$.
Since the function being trained is parity odd, taking the form of Equation~\ref{eq:parity_odd_fun},
this loss is the same whether looking at orderings $\{x, \mathrm{P}x\}$ with label real-fake or $\{\mathrm{P}x, x\}$ with label fake-real.
We can therefore make every label real--fake without consequence, and use the loss function
\begin{align}
\label{eq:logistic}
\mathcal{L}_\mathrm{which\textrm{-}is\textrm{-}real?}
&= -\frac{1}{N}\sum_{i=1}^{N}\log p(\textrm{real--fake} \mid \{x_i, \mathrm{P}x_i\}) \\
&= -\frac{1}{N}\sum_{i=1}^{N} -\log(1 + e^{-f(x)})
\label{eq:logistic_line_2}
\end{align}
for a batch of $N$ events, which follows from our choice to convert $f$ to a probability through the logistic function.
Functional forms for $f(x)$ are assigned by machine learning models through their stochastic optimizations of this loss function.

Just as classifiers end up learning divergences between data distributions,
this process learns divergences between the data distribution and its parity flipped version.
Since probabilities assigned by classifiers are invertible to likelihood ratios, classifiers approximate optimal event variables for separating their target classes, and indeed they have become standard practice for this task in high energy physics.
By the same reasoning, the parity-odd functions constructed here approximate optimal event variables for separating events from their parity-mirror images.

As an evaluation metric, we compare each learned model to the parity-symmetric hypothesis, which assigns probability $p_\textrm{sym} = 1/2$ to each ordering.
The mean log-likelihood ratio between these models is
\begin{equation}
\label{eq:quality}
Q 
= \frac{1}{N} \sum_{i=1}^{N} \log p(\textrm{real--fake} \mid \{x_i, \mathrm{P}x_i\}) 
- \log\frac{1}{2}
,
\end{equation}
which is positive only if the parity-odd model has made better predictions than the parity-even $p_\textrm{sym}$.

For perfect classification, where each event is unambiguously distinguished from its parity-flipped counterpart, $p(\textrm{real--fake} \mid \{x_i, \mathrm{P}x_i\}) = 1$ for all events, 
and $Q$ has a maximum of $\log 2 \approx 0.693$.
If both asymmetric and symmetric models predict equally well, then $Q$ takes a value of $0$,
and $Q$ is not bounded from below.

A positive $Q$ with sufficiently large $N$ is evidence for parity violation, since they imply a large likelihood ratio.
We also present error bars for estimates of the limiting value of $Q$ as $N\rightarrow \infty$, which are constructed in the usual way from the mean and standard deviation of its summands.

\subsection{Machine learning models}
\label{sec:learning_models}
This section details the BDT, NN and CNN machine learning algorithms we use for the results in this paper.
The dataset is split into subsets training~:~validation~:~testing in the ratio $60:20:20$. 
Initial tuning was performed on the training and validation sets, primarily aiming for sensitivity to the $\lambdaPV = 1$.
Early stopping methods are used to increase robustness to data from different models.
The testing set was not looked at until all models and data were finalized in preparation of the results presented in this paper.

For input features, both the BDT and NN models use only jet momenta from the truth-jet or reco-jet datasets
in the rotation-invariant coordinates $\vec q^{\,j_i}_a$ of Section~\ref{sec:invariant_jets}.
Where no fourth jet exists, we set its momenta to zero.
No derived features are included.
Separately, the CNN model processes only image data.

\subsubsection{Boosted decision tree}
The BDT uses
XGBoost~\cite{xgboost} in its
scikit-learn interface~\cite{scikit-learn, sklearn-api}
with the loss function of Equation~\ref{eq:logistic} implemented as described in~\cite{tombs2021which}.
Tuning found that parameters that slow the learning process were effective,
plausibly due to the subtlety of parity violation in the PV-mSME;
all BDTs are trained with a learning rate of $0.1$, with 
\lstinline{n_estimators=1000},
\lstinline{min_child_weight=10000},
and
\lstinline{tree_method="hist"}.

To implement early stopping, 
we evaluate every $50\textrm{th}$ iteration of the BDT on validation data, 
and choose the iteration with the best $Q$ on the validation set to use in testing. 

\subsubsection{Neural network} 
The NN is a multilayer perceptron with ReLU activation functions, 
three hidden layers of widths $(100, 100, 10)$,
and $50\%$ dropout between the second and third hidden layers.
It is implemented using the Haiku~\cite{haiku2020github} neural network library in JAX~\cite{jax2018github}
and optimized with Adam~\cite{adam}
(default settings and learning rate $0.001$)
implemented in Optax~\cite{optax2020github}.
Network inputs are pre-scaled by a mean and standard deviation in each coordinate.

Training is performed in steps with $10\,000$ examples per batch,
and is evaluated on a validation set after each $1000\textrm{th}$ step.
Training is terminated when the validation score does not increase for $10$ consecutive rounds of evaluation, up to a maximum of $100$ rounds in total.

\subsubsection{Convolutional neural network}
The CNN is implemented using PyTorch~\cite{pytorch}. 
It is trained with Adam~\cite{adam} using a learning rate of $0.001$ and L2 regularization penalty of $0.1$. 
The CNN design consists of two convolutional layers with $5\times5$ kernels, each outputting 6 channels.
These are followed by two fully connected layers of widths $96$ and $10$.
The Leaky~ReLU activation function is used, with a negative slope of $0.01$, to generate non-linearity between layers.
Invariance to rotations are built into the CNN design as described in Section~\ref{sec:invariant_images}.

We again use early stopping to train on batches of $512$ images until the validation score saturates, The validation score $Q$ is evaluated every $1000\textrm{th}$ step, and termination occurs when it has not increased for $10$ consecutive evaluation rounds.

\FloatBarrier
\section{Results}
\label{sec:results}
\begin{figure}[tp]
\centering
\includegraphics[width=0.999\textwidth]{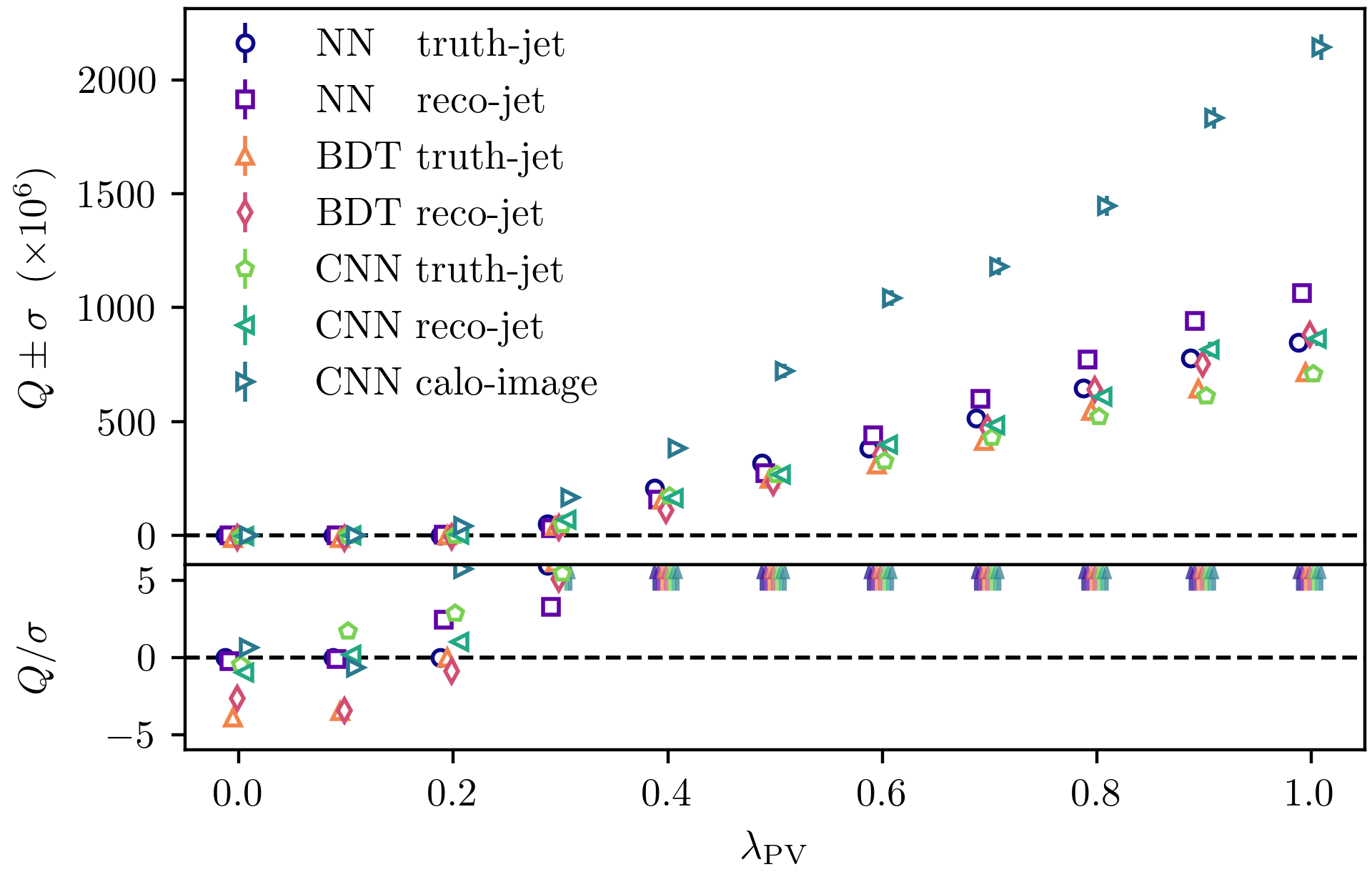}
\caption{
Test set evaluation score $Q$ as a function of the coupling $\lambdaPV$, displayed for BDT, NN and CNN models applied to truth-jet, reco-jet and calorimeter-image datasets.
A new model is trained and tested for each value of $\lambdaPV$.
All points use approximately two million events in their testing sets.
Markers at each $\lambdaPV$ point are slightly offset to aid visibility.
}
\label{fig:plots:quality_both}
\end{figure}

Test results displayed in Figure~\ref{fig:plots:quality_both} show positive $Q$ values, which demonstrate sensitivity of our method to parity violation in simulated PV-mSME data for $\lambdaPV \approx 0.3$ and above.
All models and data representations are effective, but some are more effective than others.
As might be hoped, $Q$ are seen to increase with $\lambdaPV$.

\begin{figure}[t]
\centering
\includegraphics[width=0.999\textwidth]{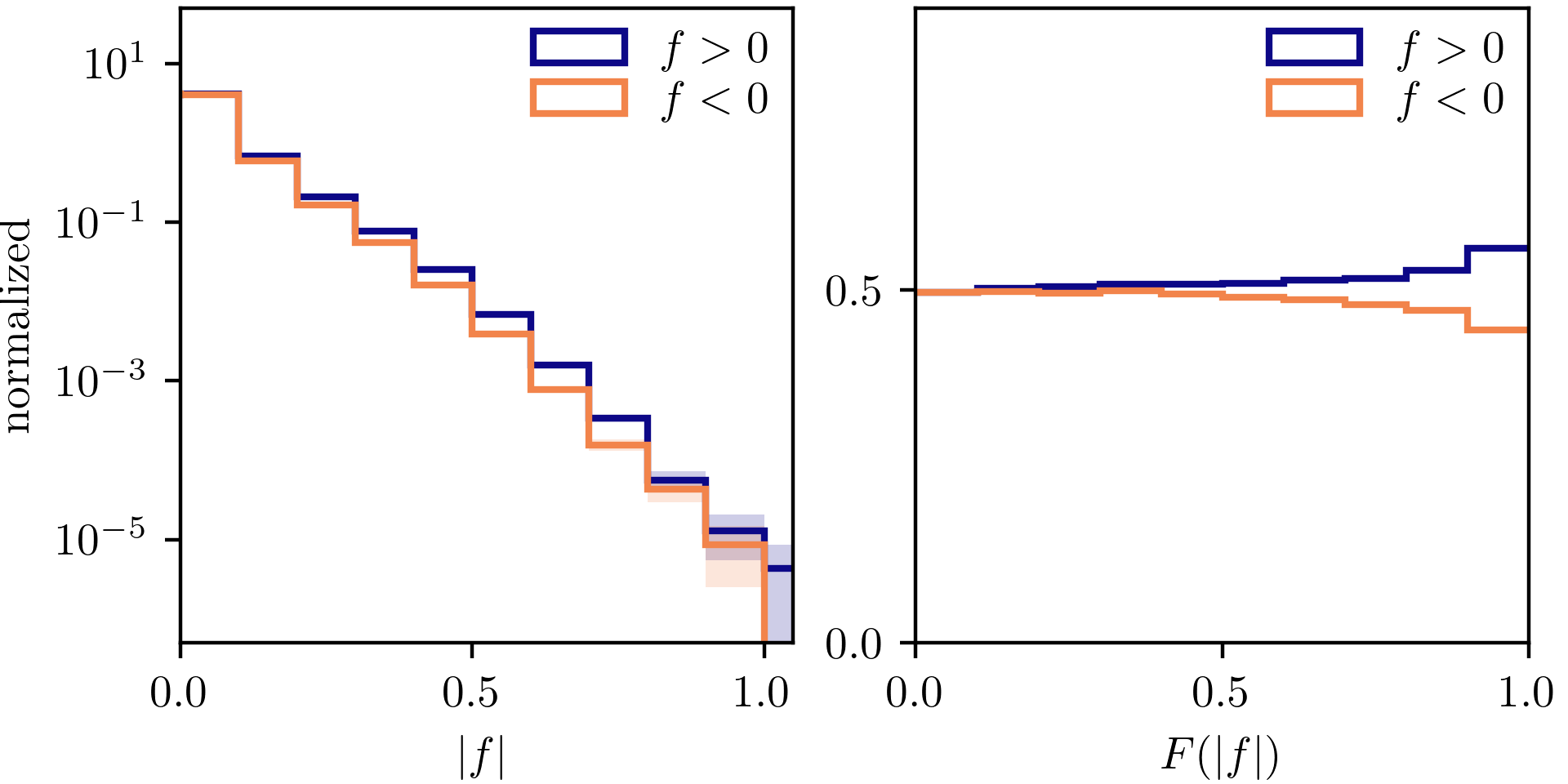}
\caption{
Distribution of the parity-odd NN variable $f$: (left) unscaled, and (right) transformed to target a uniform distribution.
The positive and negative halves of the distributions are overlaid.
Testing data are plotted with statistical error bars.
}
\label{fig:plots:hist_both_net}
\end{figure}

A representative output distribution from the NN trained on reco-jet outputs with $\lambdaPV = 1$ is shown in Figure~\ref{fig:plots:hist_both_net}. 
On the right of this plot, the variable is re-scaled to target a uniform distribution, as in Figure~\ref{fig:plots:hist_both_alpha_net_transformed_linear}. 
The clear asymmetry between the two halves of the histogram demonstrates that parity violation in this dataset is visible with the trained event variable. 
Here, the evaluation score (from Equation~\ref{eq:quality}) is $Q = (1.06 \pm 0.03) \times 10^{-3}$ with $2.3$ million testing data. 
Since the training and validation sets consume four times as many data again, this corresponds to a luminosity about of $53~ \mathrm{fb}^{-1}$ at the Standard Model cross-section of $0.22~\mathrm{nb}$.

It can be helpful to understand some of what a classifier is doing. 
For this NN model and dataset, visualizations of what has been learnt in kinematic space are given in Appendix~\ref{app:output_visualization}.

Despite the considerable noise from the various phases of physics and detector simulation,
there is a trend for models to perform \emph{better} on reco-jets than truth-jets.
Not only is the method robust to simulation and reconstruction noise, but it also appears 
to be able to scrape additional information from that noise!

Flavour information may play a significant part;
flavour is hidden from the truth-jet event variables, but does affect showering and hadronization into reco-jets. 
Different characteristics between gluons and the quark flavours may leave traces in the distributions of reconstructed jets, which the algorithms see. 
Further study of flavour information in truth data, presented in Appendix~\ref{app:truth-information},
finds that flavour information greatly increases sensitivity to parity violation.

Similarly, effects due to colour connections between partons may also play a role.
The strikingly stronger performance for the CNN with calo-images is consistent with these ideas, as it preserves some more details than clustered jets.
The CNN working with calo-images additionally accesses softer and more numerous jets than the other representations, which only have jets selected above $\pt$ thresholds.

However, this discussion of information content may be too hasty;
since all results depend entirely on the successes of learning algorithms, all differences can also be explained by how well those algorithms have performed.
That performance is influenced both by algorithm design and tuning (on the training and validation sets), which was conducted manually by us as users.

\section{Discussion and conclusions}

We have demonstrated a method to perform model-independent searches for parity-violating non-Standard-Model physics in LHC data.
This method works by training parity-odd event variables on one dataset, and evaluating them on another. 
Various technologies may be used to implement these parity-odd functions --- BDTs, NNs, and CNNs are all demonstrated to be effective.

Previous work~\cite{lester2021stressed, tombs2021which} has been extended with closer approximations to real particle physics data by simulating collider events in a parity violating quantum field theory,
with reconstruction from a detector simulation and approximated triggering.
We see from Figure~\ref{fig:plots:hist_both_alpha_net_transformed_linear} that this method achieves greater sensitivity than the previous search for parity violating physics at the LHC~\cite{Lester:2019bso}.

Parity violation may manifest itself in unexpected and unforeseen ways in nature, and in ways that are detectable at the LHC.
Indeed, parity-violating signals may already have been produced, but they have not yet been sought;
to find parity violation, it is first necessary to search.
The methods developed in this paper shrink a boundless space of previously unexplored parity violating models to a manageable size.

\FloatBarrier
\section*{Acknowledgements}

We thank members of both the Cambridge ATLAS and Pheno Working groups for valuable feedback and for enduring our endless mutterings about parity and tests for symmetry violations.
In particular, DN would like to thank Tina Potter.
Furthermore, without Olivier Mattelaer's valued assistance on matters relating to MadGraph the paper would have been impossible in its present form. 
The support of Peterhouse in hosting workshops associated with this paper is also acknowledged. 
RT, DN and CGL are all supported by the Science and Technology Facilities Research Council.  

\bibliography{bib}{}
\bibliographystyle{unsrturl}

\FloatBarrier
\appendix

\section{Additional content}

\subsection{Interpretations}
\label{app:interpretations}
\begin{figure}[tp]
\centering
\includegraphics{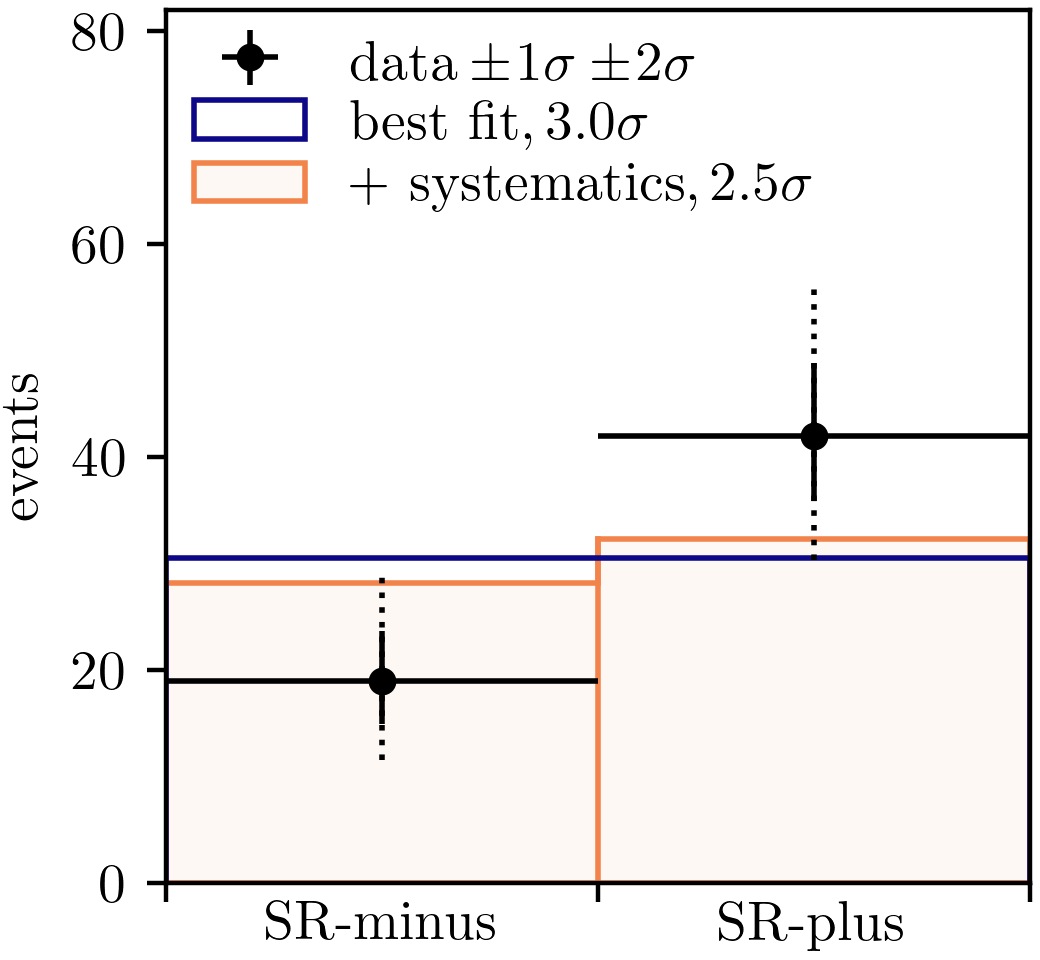}
\caption{
Illustrated analysis of parity-opposite histograms in a minimal two-bin example.
A parity-asymmetric model assigns separate expectations in each bin, so is able to fit both perfectly.
The purple line shows the best fit from a parity-symmetric model, for which both expectations must be equal.
The orange histogram modifies the parity-symmetric model with constrained `systematic' variations, which can allow small amounts of parity violation to account for known biases or uncertainties;
these variations give a slightly better fit, reducing the significance.
We assign significances $\sigma$ from maximum likelihood ratios by $\sigma^2 = -2\log [\max L_1(\theta) / \max L_2(\theta)]$, where $L_1(\theta)$ and $L_2(\theta)$ are the likelihood functions of two alternative models.
}
\label{fig:plots:two_bin_stats}
\end{figure}

The two lines in Figure~\ref{fig:plots:hist_both_net} represent histogram yields in different bins, which are separated by the parity flip operation.
An example statistical interpretation of histograms like these is illustrated in Figure~\ref{fig:plots:two_bin_stats}, which uses the common method of fitting two likelihood models; in this case, one model is one parity-even and the other is not. 
The parity even model must assign equal background expectations to both bins, up to modifications by known biases or systematic uncertainties.
The parity-odd model can assign different expectations, so fits both bins perfectly.

\subsection{Output visualization}
\label{app:output_visualization}
\begin{figure}[tp]
\centering
\begin{subfigure}[b]{\textwidth}
    \centering
    \includegraphics[width=0.999\textwidth]{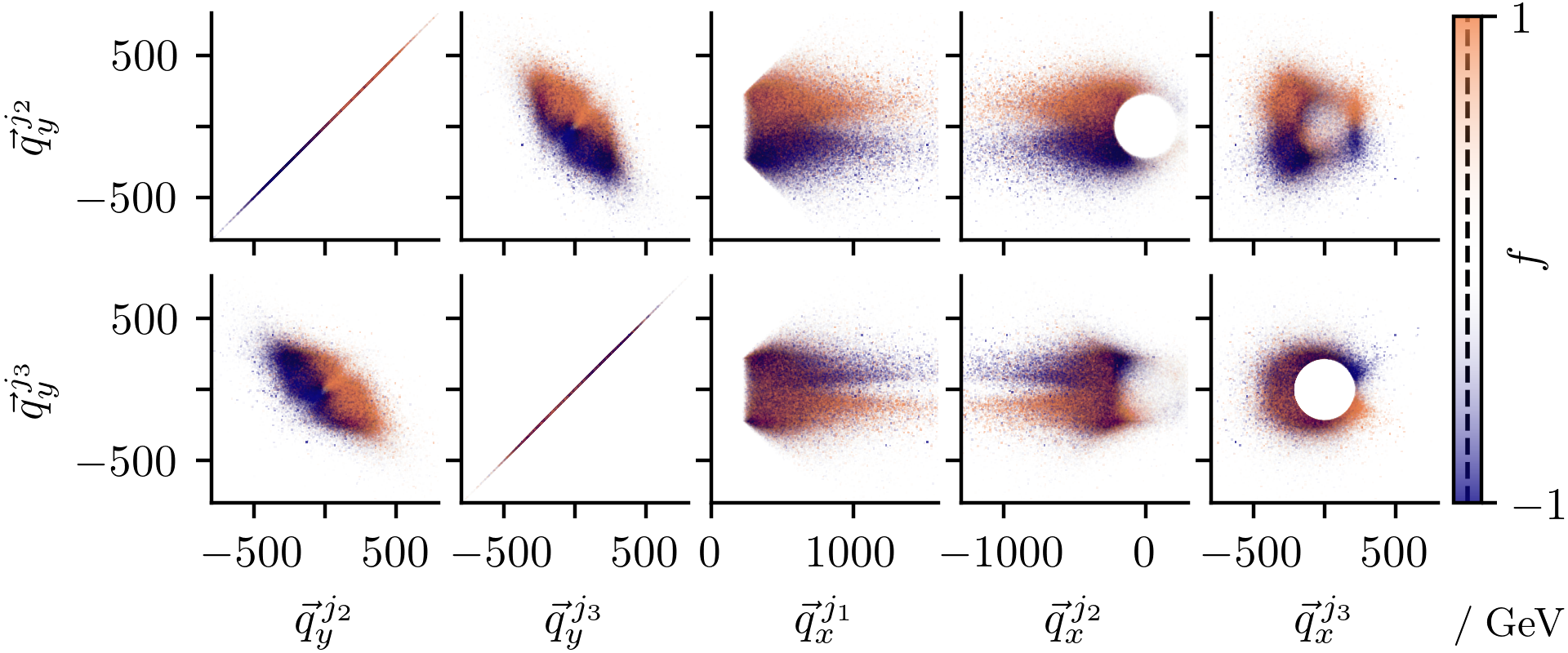}
    \caption{transverse--transverse}
\end{subfigure}
\\[1em]
\begin{subfigure}[b]{\textwidth}
    \centering
    \includegraphics[width=0.999\textwidth]{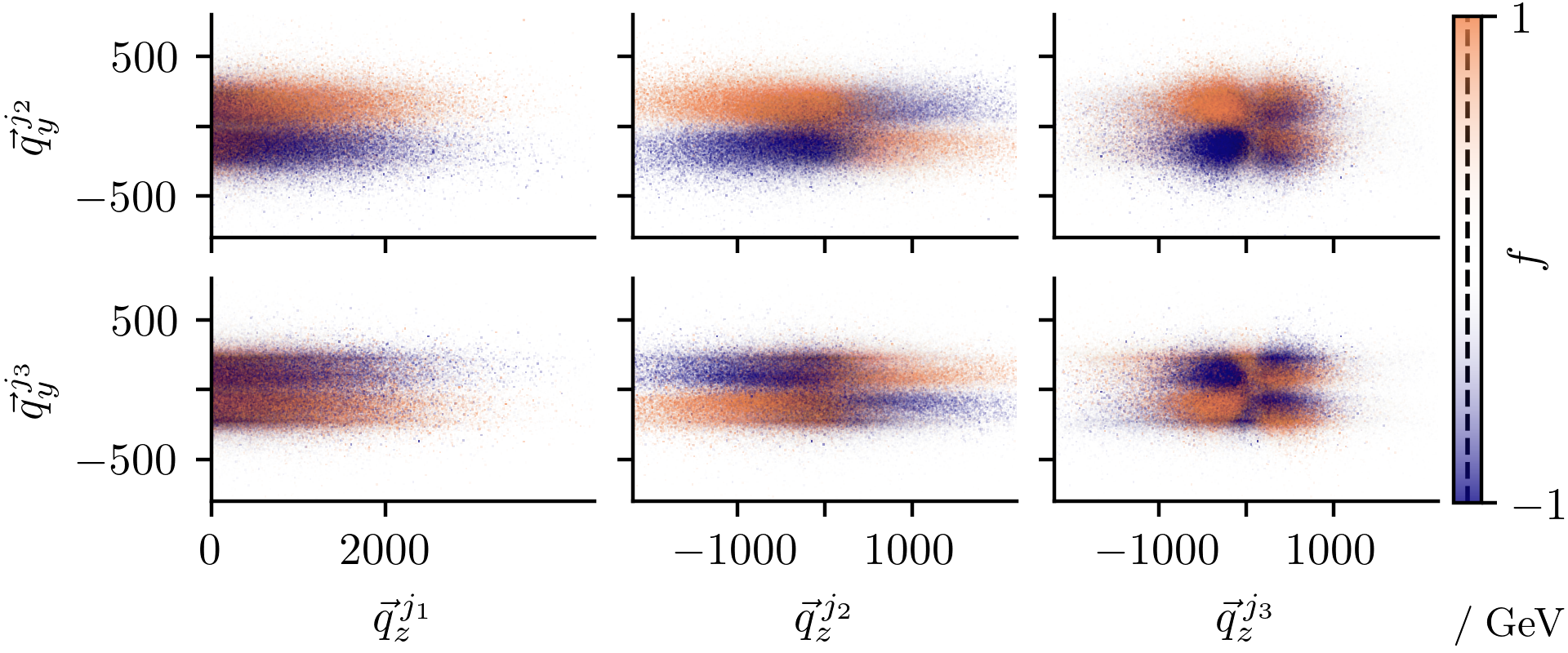}
    \caption{transverse--longitudinal}
\end{subfigure}
\caption{
Kinematic distributions of the parity-odd variable $f$ from the reco-jet NN with $\lambdaPV = 1$ in the rotation-invariant $\vec q^{\,j_i}_a$ coordinates defined in Section~\ref{sec:invariant_jets}.
The plots contain one million data from the testing dataset, scattered with colour and transparency set by the value of $f$.
Although each point has relatively low transparency, overlaid points accumulate to the observed distributions.
As seen from Figure~\ref{fig:plots:hist_both_net}, the vast majority of data have small $|f|$;
only the most parity-violating phase space is visible here.
If the NN is accurate, then colours in blue ($f < 0$) should have lower density than those in orange ($f > 0$); as seen, the difference is subtle.
}
\label{fig:plots:kin}
\end{figure}

Visualizations of what has been learned by an algorithm can be interesting and useful checks.
In this appendix, we visualize the NN parity variable trained on reco-jet data with scatter plots in which every point $x$ is coloured by its NN output $f(x)$.
These plots are shown in Figure~\ref{fig:plots:kin}.

Most events have $|f(x)| \ll 1$, as shown in Figure~\ref{fig:plots:hist_both_net},
so parity-violating phase space is highlighted by the opacity of scattered data points with $|f(x)|$.

In the rotation-invariant coordinates $\vec q_a^{\,j_i}$, described in Section~\ref{sec:invariant_jets},
only the non-zero $\hat y$ components are parity odd; all other event variables are parity even.
To show parity violation, we therefore include at least one $y$ component in each plot.
Any scatter plot in only parity-even variables cannot see parity violation and appears as a monochrome blob.

\subsection{Truth information}
\label{app:truth-information}
\begin{figure}[tp]
\centering
\includegraphics[width=0.999\textwidth]{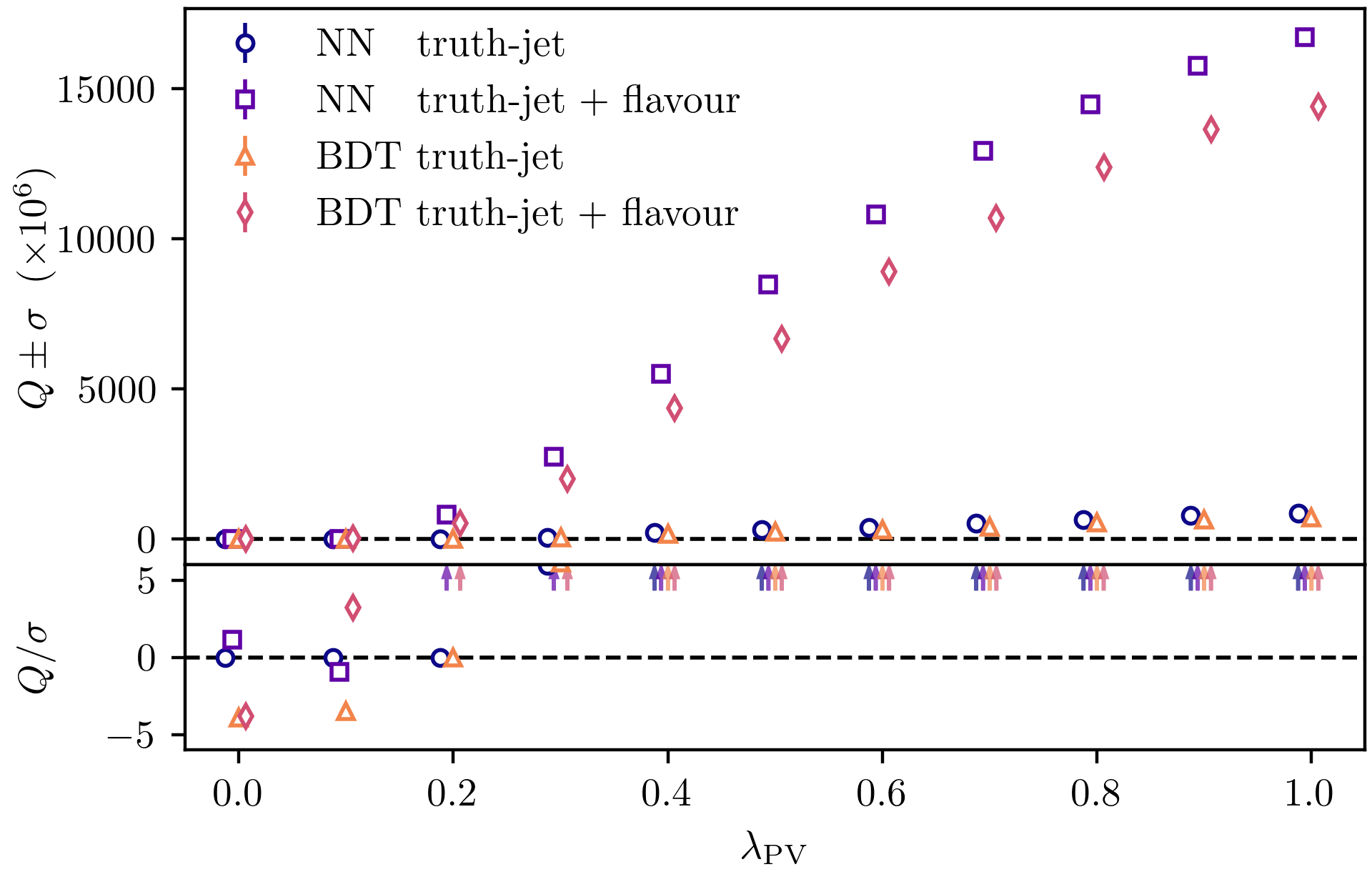}
\caption{
Test set evaluation score $Q$ as for small values of $\lambdaPV$, displayed for BDT and NN models applied to truth-jet data with and without the inclusion of parton flavour information.
Small values of $\lambdaPV$ are further investigated in Figure~\ref{fig:plots:quality_truth_net}.
All points use approximately two million events in their testing sets.
Markers at each $\lambdaPV$ point are slightly offset to aid visibility.
}
\label{fig:plots:quality_flavour}
\end{figure}

\begin{figure}[tp]
\centering
\includegraphics[width=0.999\textwidth]{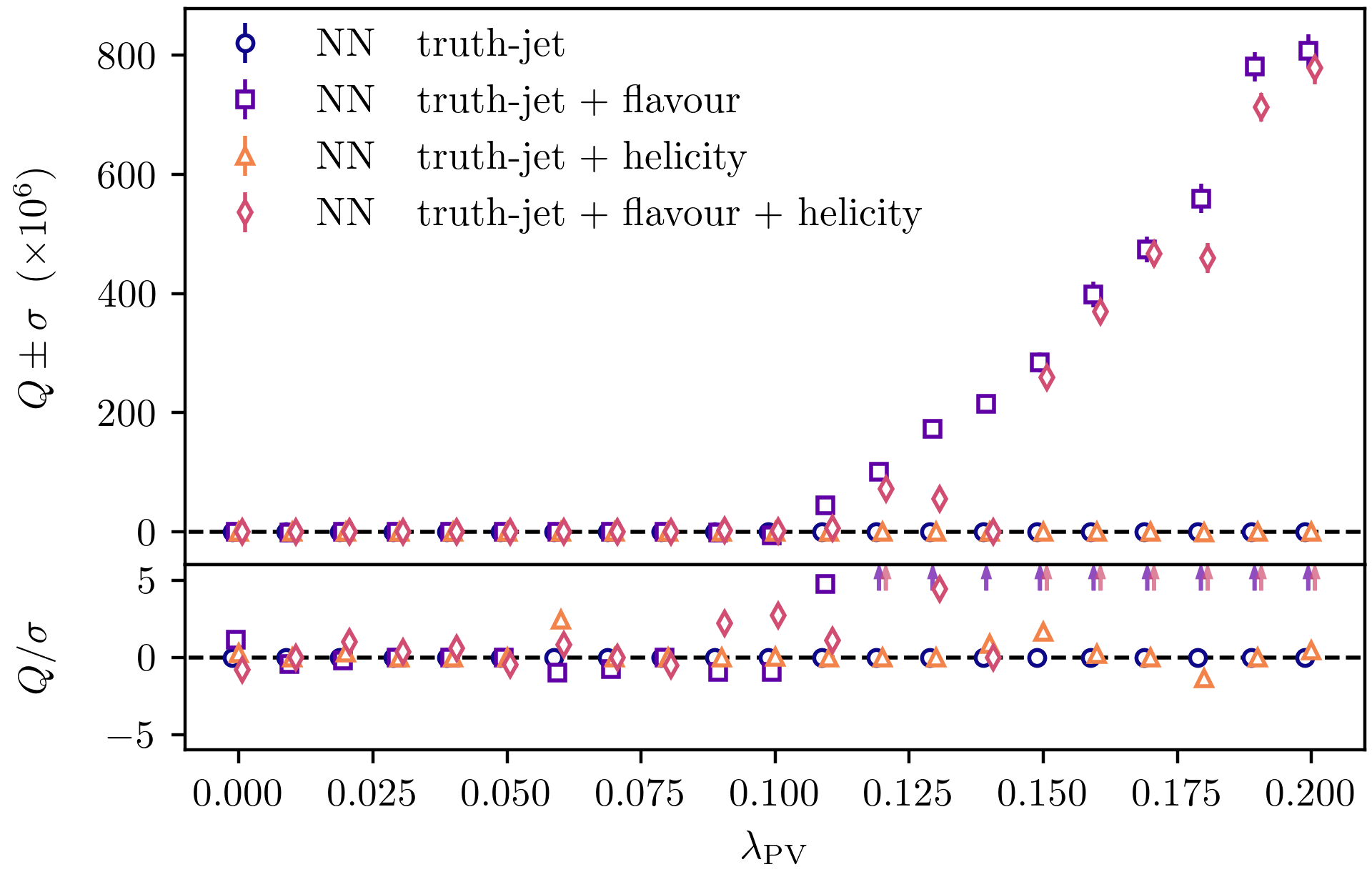}
\caption{
Test set evaluation score $Q$ for small values of the coupling $\lambdaPV$, displayed for the NN model applied to truth-jet data with and without the inclusion of parton helicity and flavour information.
This figure explores small values of $\lambdaPV$ to observe the switch-on of sensitivity
in the leftmost parts of Figure~\ref{fig:plots:quality_flavour}.
All points use approximately two million events in their testing sets.
The algorithm has performed poorly for some points; it is not perfect!
Markers at each $\lambdaPV$ point are slightly offset to aid visibility.
}
\label{fig:plots:quality_truth_net}
\end{figure}

Although parton flavour and helicity are features of the simulations, they are not included in the truth-jet dataset.
Flavour and helicity are, however, parity-even event variables which could be approximately `tagged' for with data in appropriate detectors.
This appendix investigates the effects of including flavour and helicity as additional features in variable training.
Flavour is encoded with the PDG Monte Carlo Particle Numbering Scheme~\cite{pdg},
and helicity is encoded by $\pm1$,
which are both pre-scaled by a mean and standard deviation for the NN.

Flavour information drastically boosts sensitivity, as demonstrated in Figure~\ref{fig:plots:quality_flavour} for the same $\lambdaPV=0\ldots 1$ scan as used Figure~\ref{fig:plots:quality_both}.

To investigate both the switch-on in sensitivity and the joint effect with helicity, a narrower scan is performed in Figure~\ref{fig:plots:quality_truth_net} for $\lambdaPV=0\ldots 0.2$.
Helicity appears to not help, and in fact hurts performance, as might be expected from the inclusion of irrelevant information.

\subsection{Rotated PV-mSME}
\label{app:rotation}
The use of a constant coupling matrix $(c_{A})_{\mu\nu}$  in Section~\ref{sec:pv_physics}  means that the main PV-mSME results of the paper
assume either: 
(i) that the detector does not rotate in space; 
or (ii) that the detector rotates with the earth, but the data being analysed are only those which were recorded when the detector had a particular orientation in space; 
or (iii) that the frame in which the Lorentz-violating effects are constant is `earth centric' or `dragged around' by the earth. 
The first is not physical, the second would decimate the  cross-section available for analysis, while the third is unlikely on physical grounds.  
Despite those drawbacks, we justify the decision to use a constant coupling matrix $(c_{A})_{\mu\nu}$ in Section~\ref{sec:pv_physics} on the grounds that it allows a proof-of-principle to be demonstrated in the simplest manner with the fewest distractions.

However: since the earth does rotate, it is interesting to ask what magnitude of degradation  would be seen in our results if the earth's rotation were allowed to induce a corresponding time-dependence into the couplings
$(c_{A})_{\mu\nu}$ in the detector frame. Practically, this means adding the orientation of the detector to each event in the data so that the learning algorithms can learn about alignment-dependent parity violation.
If, for example, the data were only parity violating in alignments close to a special axis, the algorithms
could learn that and automatically implement an effective selection by mapping all other data to $0$ in 
their parity-odd output.
And they could learn about other alignment-dependent parity-violating effects if they existed.

To demonstrate this approach, we imagine a simple case of an East-West aligned detector (as ATLAS is, approximately),
and consider daily rotations of a planet which is otherwise floating stationarily in space 
(that is, not otherwise orbiting a star about a different axis).
Since our analysis is invariant to rotations in $\phi$ (about the beam axis), this daily rotation has an effect equivalent to rotating the
detector about a single transverse axis (by an angle $\theta$), 
independent of latitude.\footnote{The latitude of Geneva is therefore not an input to the analysis, and no generality is lost as a consequence. Put another way: any approximately east-west-aligned and $\phi$-symmetric detector may be assumed to be near 
the North Pole, without loss of generality.}

To study this example, we simulate truth-jet datasets from the PV-mSME with its coupling matrix rotated 
for $24$ different earth-rotation angles $\theta$ spaced evenly in $[0, 2\pi)$,%
\footnote{%
Continuous rotations would be more realistic, but we opt for this discrete approximation to avoid 
substantial changes to the simulation software.
}
such that an angle of $0$ recovers the PV-mSME.
In addition to the invariant jet momentum features described in Section~\ref{sec:invariant_jets}
we also encode the $\theta$ rotation of each event with the two features $\sin\theta$ and $\cos\theta$.%
\footnote{%
This $(\sin\theta, \cos\theta)$ pair-encoding avoids the discontinuities of $\theta$ 
(at $0$ or $2\pi$), and relates linearly to dot products with directions.
Empirically, it appeared to improve results in preliminary tests on the validation set.%
}
Training, validation, and testing sets are prepared by subsampling and shuffling portions of these data;
we take samples from each rotation with rates proportional to their production cross-sections, 
which we calculate from MadGraph and the efficiency of our kinematic selections.

Cross-sections in the rotated PV-mSME vary hugely with $\theta$.
(Although this anisotropy is a clear sign of new physics, it is not a direct sign of parity violation.)
These variations are approximately sinusoidal, with minima at $0$ and $\pi$,
maxima at $\pi/2$ and $3\pi/2$, and a maximum-over-minimum ratio of $13.5$.
Since parity violation is strongest at $0$ and $\pi$ (as designed in Appendix~\ref{app:pvsme_details}),
this drastically dilutes the observable parity violation.
When training our standard learning models (described in Section~\ref{sec:learning_models}) on these rotated data, they report $Q \leq 0$ in validation, so not find parity violation.

After tuning new model designs towards these different data, however, 
we do find some evidence for parity violation in these rotated samples.
With the diluted parity violation in the rotated sample, the standard NN tended to collapse
towards all-$0$ outputs;
this error was avoided by simplifying its design to three hidden layers of widths $(20, 20, 10)$,
and no dropout layer.
For statistically significant results, we also triple the data size from $10$~million 
(as stated in Section~\ref{app:simulation_details}) to $30$~million, 
with the same training~:~validation~:~testing split of $60:20:20$.

After finalizing this tuned NN, we find $Q = (7.6 \pm 2.6) \times 10^{-6}$
in the testing set, which is a positive indication of parity violation.
This result is weaker than our other examples of evidence for parity violation,
such as the results displayed in Figure~\ref{fig:plots:quality_both}, but 
still corresponds to a substantial $\log$ likelihood ratio of $45.6$ in favour 
of the parity violating NN model.

\section{Simulation details}
\label{app:simulation_details}
We simulate parton scattering at leading order using MadGraph5\_aMC@NLO v3.3.0~\cite{MadGraph}.
By default, MadGraph evaluates its matrix elements in the centre of mass frame;
since the mSME is not Lorentz invariant, we modify code generated by MadGraph
to evaluate its matrix elements in the lab frame instead.
We use a custom implementation of the PV-mSME quark-quark-gluon vertices in a model imported into MadGraph.
The simulated hard scatter is of LHC-like proton-proton collisions at $\sqrt{s} = 13~\eV[T]$ 
scattering to three or four gluons or light quarks 
(up, down, strange, or charm) each with
$\pt > 200~\eV[G]$ and $\abseta < 3.2$, and selected for CKKW-L merging with $\ktdurham = 200~\eV[G]$~\cite{ckkw}.
Other MadGraph parameters of are left at their default values ---
in particular, partonic jets must be separated by $\Delta R_{jj} > 0.4$,
and we use the NNPDF2.3LO1 PDF set~\cite{nnpdf2013}. 

The parton shower and hadronization are simulated with \PYTHIA~8.235~\cite{pythia}. Detector reconstruction is performed with the Delphes~3.5.0~\cite{de2014delphes} approximation to the ATLAS detector.
The output from Delphes includes anti-$\kt$ jets~\cite{antikt} at $\Delta R = 0.4$, and energy deposits in the calorimeters.
Delphes performs jet clustering with the FastJet package~\cite{Cacciari:2011ma,Cacciari:2005hq}.

The effect of pileup is simulated in Delphes by overlaying a mean of 50 minimum-bias events to each event.
Delphes sub-samples from a single batch of pileup events for each batch of events processed;
this raises a risk of bias if pileup events are repeated enough for our learning algorithms to recognize them individually.
We mitigate this risk by using large numbers of pileup events and not sharing any between training and testing datasets.
We generate events in batches of $200\,000$, and for each batch simulate $200\,000$ uniquely-seeded pileup events.

To approximate the efficiency plateau of a three-jet trigger, events are accepted only if they are reconstructed with at least three jets with $\pt > 220~\eV[G]$ and $\abseta < 2.8$.
Additional reconstructed jets are included only if they satisfy $\pt > 30~\eV[G]$ and $\abseta < 2.8$.

We generate $500$ batches of $200\,000$ truth events for each model specification.
These events acquire weights in their processing to reconstruction level, which we unweight by downsampling
with respect to the maximum of all weights in the simulation.
Depending on $\lambdaPV$, around $20\textrm{--}30\%$ of events survive triggering and kinematic selections, and of those accepted around $60\textrm{--}40\%$ survive unweighting.
After these reductions, the reco sets include $9.2$~million events at $\lambdaPV=0$,
and $11.5$~million events at $\lambdaPV=1$.
Truth sets have more events available since they do not suffer from unweighting, 
so approximately match reco we use exactly $10$~million truth events for each model.

\FloatBarrier
\section{PV-mSME details}
\label{app:pvsme_details}

\subsection{Notation}

Our notation is related to Equation 11 of~\cite{LIV:Main:Colladay:1998fq}, 
which defines coupling matrices $(c_X)_{\mu\nu AB}$ for $X \in \{Q, U, D\}$
with generation indices $A$ and $B$.
The PV-mSME defines these coupling matrices to be diagonal in the first two generations:
$(c_X)_{\mu\nu AB} = (c_X)_{\mu\nu}\,\mathrm{diag}(1, 1, 0)_{AB}$.
Couplings split into an axial part
$(c_{A})_{\mu\nu} = (c_{U})_{\mu\nu} - (c_{Q})_{\mu\nu}$,
and a vector part which we define to vanish
$(c_{V})_{\mu\nu} = (c_{D})_{\mu\nu} + (c_{Q})_{\mu\nu} = 0$
The PV-mSME therefore couples quarks within the first two generations only, 
has the same couplings within each of those generations, and does not mix between them.

After some rearrangement, the quark-quark-gluon Feynman Rule of Equation~\ref{eq:pv-sme-qqg-vertex} 
can be read from the Lagrangian in these terms.

As stated in~\cite{LIV:Main:Colladay:1998fq}, 
the full coupling matrices must be Hermitian in their generation indices and traceless in their Lorentz indices.
Since our couplings are on the generation diagonal, they must be real to satisfy the Hermitivity constraint.

\subsection{Couplings}

Within those real and traceless constraints, the specific form of the
PV-mSME coupling matrix $(c_{A})_{\mu\nu}$ of Equation~\ref{eq:camunu_pv_mSME}
is chosen primarily from empirical results of numerical experiments.
However, it has some plausible justifications, which we develop here.

We seek couplings which not only generate parity violation,
but which make it visible when blind to rotations by $\phi$ about the beam axis 
and by $180^\circ$ about a perpendicular axis (to swap the beams).
Variation of a differential cross-section under these rotations is of no benefit,
since the rotation-invariant observer sees only an average.
Variations may plausibly, however, cause parity violation to be `washed out' from a rotation-invariant perspective.
We therefore attempt to choose couplings which minimize the dependence of cross-sections on these rotations.

The PV-mSME introduces terms into matrix elements that do not appear in the Standard Model alone. 
Among these non-standard terms are expressions of the form $(c_{A})_{\mu\nu}p^\mu q^\nu$,
which we abbreviate here as $\xi = p^T C q$.
Parameterizing axial and beam-swap rotations with the matrix
\begin{equation}
S(\phi, \pm)
= 
\begin{pmatrix}
    1 & 0 & 0 & 0 \\
    0 & \cos\phi & \mp\sin\phi & 0 \\
    0 & \sin\phi & \pm\cos\phi & 0 \\
    0 & 0 & 0 & \pm1 \\
\end{pmatrix},
\end{equation}
$\xi$ may be seen to transform to $\xi(\phi, \pm) = p^T S(\phi, \pm)^T C S(\phi, \pm) q$.
If one were to require that $\xi(\phi, \pm)$ be independent of $\phi$ for all $p$ and $q$, 
then one would need to take $C$ to be of the form
\begin{equation}
C
= 
\begin{pmatrix}
    c_{00} & 0 & 0 & c_{03} \\
    0 & c_{22} & -c_{21} & 0 \\
    0 & c_{21} & c_{22} & 0 \\
    c_{30} & 0 & 0 & c_{33} \\
\end{pmatrix}
.
\end{equation}
If one were to subsequently require that $\xi(\phi, \pm)$ be proportional to $\pm$ for all $p$ and $q$, 
then the diagonal elements of $C$ would be set to zero, resulting in:
\begin{equation}
\label{eq:xi_fourterms}
\xi(\phi, \pm)
=
\pm(p_0 q_3 c_{03} + p_3 q_0 c_{30} + p_2 q_1 c_{21} - p_1 q_2 c_{21})
~.
\end{equation}
This choice (fixed magnitude but not fixed sign) does not lead matrix elements to be invariant under beam swaps, but does give opportunities for them to have reduced dependence on this operation,
perhaps arising from even powers or cancellations against sign-flipped terms.

Pleasingly, the form shown in \eqref{eq:xi_fourterms} contains both 
parity even ($p_1 q_2$, $p_2 q_1$) and 
parity odd ($p_0 q_3$, $p_3 q_0$) terms.
Violation of parity requires interference between parity-odd and parity-even terms;
parity odd terms alone are not sufficient, since cross-sections are proportional to the modulus of the matrix element squared.

Requiring $\xi(\phi, \pm)$ to be parity asymmetric (for some $p$ and $q$)
therefore forces $c_{21}$ and at least one of $c_{30}$ and $c_{03}$ to be non-zero.
Without loss of generality we can therefore choose $c_{21} = 1$, absorbing any overall scale into $\lambdaPV$, 
and are left with the freedom to choose values for $c_{30}$ and $c_{03}$.

In summary, requiring $\xi(\phi, \pm)$ to be
independent of $\phi$, 
invariant in magnitude to $\pm$ flips,
and parity asymmetric implies
\begin{equation}
C
= 
\begin{pmatrix}
    0 & 0 & 0 & c_{03} \\
    0 & 0 & -1 & 0 \\
    0 & 1 & 0 & 0 \\
    c_{30} & 0 & 0 & 0 \\
\end{pmatrix}
,
\end{equation}
where at least one of $c_{30}$ and $c_{03}$ is non-zero.

\begin{figure}[tp]
\centering
\includegraphics{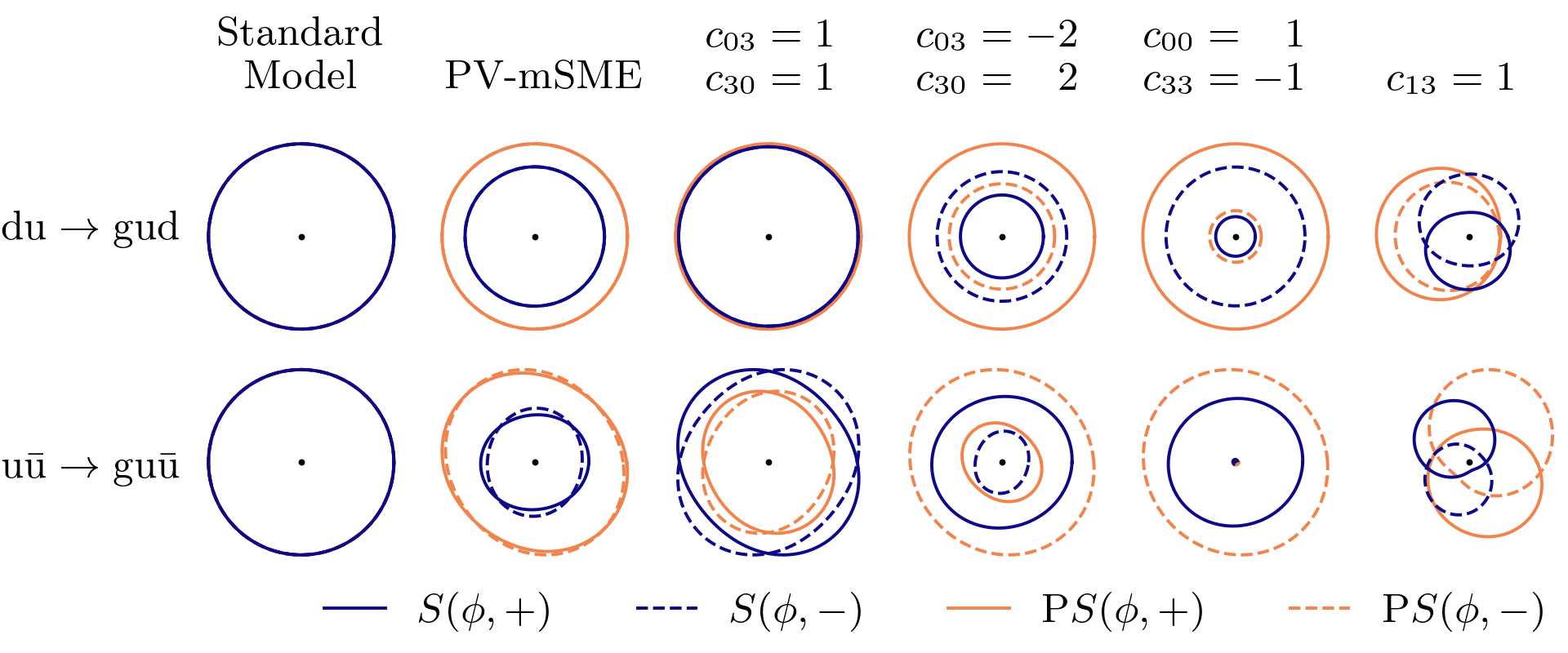}
\caption{
Helicity-summed, magnitude-squared matrix elements $|\mathcal{M}|^2$ displayed as radial distances, such that differences between the orange and blue lines indicate parity violation.
Momenta are rotated by $S(\phi, \pm)$, where $\phi$ is the angle from the vertical axis.
Parton distribution functions are invariant under these rotations, so $|\mathcal{M}|^2$ is proportional to the differential cross-section.
The Standard Model is invariant under rotation and parity operations.
The PV-mSME has $(c_{A})_{\mu\nu}$ couplings from Equation~\ref{eq:camunu_pv_mSME};
latter columns modify the $(c_{A})_{\mu\nu}$ coupling matrix by their stated assignments.
Top: 
$\vec p_\mathrm{g} = (200, 400, -600)$, 
$\vec p_\mathrm{u} = (-200, -150, 200)$,
and $\vec p_\mathrm{d} = (0, -250, 1200)~\eV[G]$.
Bottom: 
$\vec p_\mathrm{g} = (100, 250, 250)$, 
$\vec p_\mathrm{u} = (200, 250, 400)$,
and $\vec p_\mathrm{\bar u} = (-300, -500, -300)~\eV[G]$.
}
\label{fig:plots:rings_combo}
\end{figure}

The PV-mSME choice is $c_{30} = 1$ and $c_{03} = -1$.
This empirically generates visible parity violation,
and has differential cross-sections which are close to rotation-invariant when explored from individual points in phase space.
Numerical matrix elements calculated with MadGraph are illustrated at example phase points in Figure~\ref{fig:plots:rings_combo},
along with some effects from differently assigned couplings.

In the top row of Figure~\ref{fig:plots:rings_combo}, Standard Model and PV-mSME rings appear to be rotation-invariant, but they are not exactly.
On closer inspection, the PV-mSME shows sinusoidal variations of a few parts per million, 
whereas the Standard Model varies only by numerical rounding errors in parts per $10^{15}$.

\end{document}